\documentclass[10pt,journal]{IEEEtran}
\usepackage{tabularx}
\usepackage[T1]{fontenc}% optional T1 font encoding
\usepackage{algorithm}
\usepackage[noend]{algpseudocode}
\usepackage{graphicx}
\usepackage{amsmath,amssymb,amsfonts,stfloats}
\usepackage{array}
\usepackage{textcomp}
\usepackage[font=small]{caption}
\usepackage{paralist}
\usepackage{lettrine}
\usepackage{cite}
\usepackage{xcolor}
\usepackage{cancel}
\usepackage{booktabs}
\usepackage[export]{adjustbox}
\usepackage[caption=false,font=footnotesize]{subfig}
\usepackage{url}
\usepackage{acronym}
\usepackage{multicol}

\begin{document}

\title{Motion Estimation and Compensation in Automotive MIMO SAR}
%\title{Analysis and Compensation of Motion Estimation Errors in Automotive MIMO SAR}
%
%
% author names and IEEE memberships
% note positions of commas and nonbreaking spaces ( ~ ) LaTeX will not break
% a structure at a ~ so this keeps an author's name from being broken across
% two lines.
% use \thanks{} to gain access to the first footnote area
% a separate \thanks must be used for each paragraph as LaTeX2e's \thanks
% was not built to handle multiple paragraphs
%

\author{Marco Manzoni,
        Dario Tagliaferri,
        Marco Rizzi,
        Stefano Tebaldini,
        Andrea Virgilio Monti Guarnieri,
        \\ Claudio Maria Prati,
        Monica Nicoli,
        Ivan Russo,
        Sergi Duque,
        Christian Mazzucco,
        Umberto Spagnolini}
\maketitle
%
% As a general rule, do not put math, special symbols or citations
% in the abstract or keywords.
\begin{abstract}
With the advent of self-driving vehicles, autonomous driving systems will have to rely on a vast number of heterogeneous sensors to perform dynamic perception of the surrounding environment. Synthetic Aperture Radar (SAR) systems increase the resolution of conventional mass-market radars by exploiting the vehicle's ego-motion, requiring a very accurate knowledge of the trajectory, usually not compatible with automotive-grade navigation systems. In this regard, this paper deals with the analysis, estimation and compensation of trajectory estimation errors in automotive SAR systems, proposing a complete residual motion estimation and compensation workflow. We start by defining the geometry of the acquisition and the basic processing steps of Multiple-Input Multiple-Output (MIMO) SAR systems. Then, we analytically derive the effects of typical motion errors in automotive SAR imaging. Based on the derived models, the procedure is detailed, outlining the guidelines for its practical implementation. We show the effectiveness of the proposed technique by means of experimental data gathered by a 77 GHz radar mounted in a forward looking configuration.

\end{abstract}

% Note that keywords are not normally used for peerreview papers.
\begin{IEEEkeywords}
SAR, Automotive, MIMO, Autofocus, Motion compensation.
\end{IEEEkeywords}

% For peer review papers, you can put extra information on the cover
% page as needed:
% \ifCLASSOPTIONpeerreview
% \begin{center} \bfseries EDICS Category: 3-BBND \end{center}
% \fi
%
% For peerreview papers, this IEEEtran command inserts a page break and
% creates the second title. It will be ignored for other modes.
\IEEEpeerreviewmaketitle

\section{Introduction}
% The very first letter is a 2 line initial drop letter followed
% by the rest of the first word in caps.
% 
% form to use if the first word consists of a single letter:
% \IEEEPARstart{A}{demo} file is ....
% 
% form to use if you need the single drop letter followed by
% normal text (unknown if ever used by the IEEE):
% \IEEEPARstart{A}{}demo file is ....
% 
% Some journals put the first two words in caps:
% \IEEEPARstart{T}{his demo} file is ....
% 
% Here we have the typical use of a "T" for an initial drop letter
% and "HIS" in caps to complete the first word.

%%%%%%%%%%%%%%%%%%%%% Introduzione %%%%%%%%%%%%%%%%%%%
The evolution to fully-autonomous vehicles requires the usage of a huge and heterogeneous set of sensors, such as cameras, lidars, radars, acoustic, etc., to enable advanced environmental perception \cite{Marti2019ADAS_sensors}. Cameras and lidars are, respectively, passive and active optical sensors able to create high-resolution images and/or point clouds of the surrounding. If properly integrated, they can provide the vehicles with the capability of detect and classify objects in the environment. Automotive-legacy Multiple-Input Multiple-Output (MIMO) radars working in W-band ($76-81$ GHz~\cite{Hasch2012}) are widely employed to obtain measurements of radial distance, velocity and angular position of remote targets \cite{feng_lane_2019}. Advantages of radars are more than a few: they work in any weather condition, do not need any external source of illumination and are available at low cost. However, mass-market automotive radars are characterized by a poor trade-off between angular resolution - typically above $ 1$ deg -, range, bandwidth and Field Of View (FOV), challenging their usage for high-resolution environment mapping in automated driving~\cite{Hasch2012,Brisken2018}.  

%%%%%%%%%%%%%%%%%%%%% SAR automotive %%%%%%%%%%%%%%%%%%%
Significant effort was spent in recent works to increase the accuracy of environmental perception by means of Synthetic Aperture Radar (SAR) techniques~\cite{iqbal_sar_2015,feger_experimental_2017,stanko_millimeter_2016,wang_auxiliary_2019}. With SAR, a moving radar sensor is employed to synthesize a large antenna array (synthetic aperture) by coherently combining different acquisitions in different positions of the trajectory; the range resolution, dictated by the bandwidth, remains the same as for conventional real aperture radars, while the angular resolution increases proportionally to the length of the synthetic aperture (typically $\ll 1$ deg). In~\cite{iqbal_sar_2015}, an automotive SAR system has been simulated using a radar mounted on a sliding rail. In~\cite{feger_experimental_2017}, a $77$ GHz radar with $1$ GHz of bandwidth was mounted on the rooftop of a car to obtain images with resolution as small as $15$ cm. The system proved to be capable of imaging the scene composed by cars, fences, sidewalks, houses and more. In~\cite{wang_auxiliary_2019}, SAR images were used to search for free parking areas, while a $300$ GHz SAR implementation (with $40$ GHz of bandwidth) is presented in~\cite{stanko_millimeter_2016}, showing millimeter-accurate imaging capabilities on a slowly travelling van along a linear path. A preliminary investigation on a cooperative SAR system aimed at increasing the resolution in scarce bandwidth conditions is by our previous work \cite{tagliaferri_cooperative_2021}.

%%%%%%%%%%%%%%%%%%%%% Conoscenza traiettoria %%%%%%%%%%%%%%%%%%%
All the aforementioned works underline how the knowledge about instantaneous radar position is of utmost importance for automotive SAR systems. Errors in motion estimation, due to inaccurate navigation data, make the SAR images to appear rotated and defocused~\cite{Wu2012_MotionCompensation_thesis}. In principle, SAR requires navigation accuracy to be lower than the wavelength ($4$ mm in W-band) \cite{tebaldini_phase_2016}, but the requirement is on the \textit{relative} motion within the synthetic aperture, that extends up to tens of centimeters. Early works on automotive SAR~\cite{Zwick2010_motioncompensation_automotiveSAR,Zwick2011_motioncompSARgyroacc} propose simple accelerometer- and/or gyroscope-based Motion Compensation (MoCo), other ~\cite{Gisder2018_automotiveSAR_wheelspeed} consider the usage of odometric wheel speed, for approximately linear trajectories. Our previous work~\cite{tagliaferri_navigation-aided_2021} demonstrated a good SAR imaging quality in urban scenarios, employing an \textit{ad-hoc} fusion of multiple sensors such as Global Navigation Satellite System (GNSS), Inertial Measurement Units (IMUs), odometer and steering angle. However, these former works highlighted the need of a proper residual motion correction in arbitrary dynamic conditions, where automotive-grade navigation solutions are not accurate enough.

%%%%%%%%%%%%%%%%%%%%% SOA autofocus %%%%%%%%%%%%%%%%%%%
Therefore, the residual motion estimation and compensation is still and open issue. Traditionally, air-borne and space-borne SAR systems make use of radar data to refine the positioning accuracy with an \textit{autofocus} procedure~\cite{prats_comparison_2007, wahl_phase_1994, mao_two-dimensional_2016}. 
%one possible solution already utilized in air-borne and space-borne SAR applications is to exploit the radar data itself to refine the positioning accuracy. Different algorithms has been proposed in the literature, and they all go under the name of \textit{autofocus} or \textit{motion compensation} (MoCo) algorithms \cite{prats_comparison_2007, wahl_phase_1994, mao_two-dimensional_2016}.
Very little work has been done, on the other hand, in the automotive field, whose most relevant contributions on residual motion estimation and compensation are in~\cite{Zwick2009_SARforparking,kan_implementation_2020,gishkori_imaging_2021,kellner_instantaneous_2013,iqbal_imaging_2021}. In~\cite{Zwick2009_SARforparking}, for instance, an autofocus procedure has been employed for the focusing of SAR images without dealing with MIMO SAR and providing simulation results. In~\cite{kan_implementation_2020}, a complete automotive-based SAR system has been proposed, based on a new approach for motion compensation. As the vehicle changes its velocity, the radar's parameters such as the pulse repetition frequency is changed, to avoid distortions in the final SAR image. The paper however does not cope with errors in the knowledge of the vehicle trajectory. In~\cite{gishkori_imaging_2021}, the authors report an automotive SAR system, based on compressive sensing, that does not require any prior on the motion of the vehicle. However, compressive sensing approaches are known to be hardly applicable to real scenarios with non-linear and possibly high-velocity vehicle trajectories.
Two similar approaches are in \cite{kellner_instantaneous_2013} and \cite{iqbal_imaging_2021}. In the former, the authors use a standard range-Doppler radar to refine the ego-motion estimation of the vehicle. This information is not used, however, to focus a SAR image. In the latter, instead, two radars are used: one for the ego-motion estimation and the other for the SAR image formation. 
%In this contribution we developed a MoCo procedure able to use a single MIMO radar with reduced system's complexity.

%%%%%%%%%%%%%%%%%%%%% Contributi %%%%%%%%%%%%%%%%%%%
This paper proposes a complete residual motion estimation and compensation procedure for automotive SAR systems, using radar data on top of navigation ones. The technique exploits a single MIMO radar to do both a refinement of the ego-velocity and SAR image formation. Such radar can be mounted in any looking geometry (forward or side looking) opening the possibility for the formation of wide and dense SAR map of the urban environment.
We start by providing the analytical treatment of the effect of typical vehicle motion errors on SAR imaging, highlighting the major sources of image degradation and providing the theoretical requirements in terms of maximum tolerable velocity errors. Then, we outline the proposed autofocus workflow, exploiting a set of co-registered low-resolution images obtained by focusing the data received by a MIMO radar mounted on the car. The low-resolution images provide the location of a set of Ground Control Points (GCPs), that are first used to retrieve the residual Doppler frequency and, consequently, the velocity error. Finally, low-resolution images are phase-compensated and summed along the synthetic aperture, obtaining a correctly focused SAR image in any driving condition and acquisition geometry. We also provide some insights on how to properly select the GCPs, as well as on the effect of an error in the focusing height and/or angular mis-localization of a GCP on SAR imaging. The expected theoretical performance of the residual motion estimation and compensation are also assessed.  The work is validated by experimental data gathered by a 77 GHz MIMO radar mounted on the front bumper of a car, in a frontal looking configuration. The car is purposely equipped with navigation sensors to provide the a-priori trajectory estimation, input of the autofocus procedure. The results confirm the validity of the proposed approach, that allows to obtain cm-accurate images of urban environments.

%The solution proposed in the present contribution exploits a set of coregistered low-resolution images obtained by focusing the data received by a MIMO radar mounted on the vehicle. These images provide the location of a set of Ground Control Points (GCP) over which the residual Doppler frequency is estimated. From this residual Doppler frequency the motion error is derived and compensated leading to a well focused and well localized SAR image.\\
%The procedure is able to work in whatever geometry such as forward looking SAR, classical lateral-looking SAR and squinted SAR.
%This contribution also provides some insights on how to properly choose GCPs and what are the expected performances of the procedure.\\

%%%%%%%%%%%%%%%%%%%%% Organizzazione %%%%%%%%%%%%%%%%%%%
The paper is organized as follows: in Section \ref{sec:FMCW_ULA_SAR}, an introduction to Frequency Modulated Continuous Wave (FMCW) MIMO SAR processing is provided; Section \ref{sec:motion_error} reports the analytical derivation of the effects of typical motion errors on SAR focusing; Section \ref{sec:moco} describes the proposed autofocus workflow, validated with experimental data in Section \ref{sec:results}. Finally, Section \ref{sec:conclusions} draws the conclusion.

%
% needed in second column of first page if using \IEEEpubid
%\IEEEpubidadjcol
%
%%%%%%%%%%%%% GEOMETRY %%%%%%%%%%%%%%%%%%%%
\section{FMCW SAR processing}
\label{sec:FMCW_ULA_SAR}
For the sake of clarity, in this section we propose a review of the core aspects of FMCW radars \cite{meta_signal_2007}, MIMO and SAR processing. Each system is described, the geometry of the problem is explained, strengths and limitations are reported.
\subsection{FMCW Preliminaries}
%\subsection{Signal model for FMCW radars}
\label{sec:fmcw}
Let us consider a FMCW radar operating in W-band, located in the origin of a 2D scenario, emitting a chirp signal of duration $T_p$ every Pulse Repetition Interval (PRI). The emitted signal is:  
\begin{equation}
    s_{tx}(t) = \exp\{j (2 \pi f_c t + \pi K t^2)\} \times \mathrm{rect}\left[\frac{t}{T_p} \right]
\end{equation}
where $t$ is the \textit{fast-time} variable, $f_c$ is the carrier frequency, $K$ is the chirp rate measured in [Hz/s] and the overall frequency sweep covers a bandwidth $B$. After the demodulation and deramping of the received signal, the range of a target at distance $r_0$ from the radar can be estimated from the Range-Compressed (RC) datum \cite{Zaugg2015_FMCWSAR}: 
\begin{equation}
    s_{rc}(t; t_0)=T_p \,\mathrm{sinc}[T_p(f-Kt_0)]\exp\{-j2\pi f_c t_0\}
\end{equation}
where $t_0 = 2r_0/c$ is the two-way travel time of the Tx signal ($c$ is the speed of light in vacuum) and and $\mathrm{sinc}[x] = \sin x/x$.
We also assumed a fast chirp modulation where range and Doppler are decoupled.

By performing the change of variable $t = 2r/c$, we obtain:
\begin{equation}
\label{eq:range_compressed_FMCW}
    \begin{split}
        s_{rc}(r; r_0)= T_p \,\mathrm{sinc}\left[\frac{2B}{c}(r-r_0)\right]\exp\left\{-j\frac{4\pi}{\lambda} r_0\right\}
    \end{split}
\end{equation}
where $\lambda$ is the carrier wavelength. The range resolution of the FMCW radar system is therefore 
\begin{equation}
    \rho_r = \frac{c}{2B}.
\end{equation}
For an exemplary bandwidth $B=3$ GHz, the range resolution is approximately $5$ cm. A radar system with a single antenna will provide no resolution at all in the direction orthogonal to range (azimuth): every target at the same distance from the sensor will be integrated in the same resolution cell. To provide the angular resolution it is common practice to use arrays of antennas, either real or virtual. 
\subsection{MIMO Processing}
%\subsection{MIMO vs. SAR Processing}
\label{sec:ULA}

\begin{figure}[!t]
    \centering
    \subfloat[][MIMO]{\includegraphics[width=0.7\columnwidth]{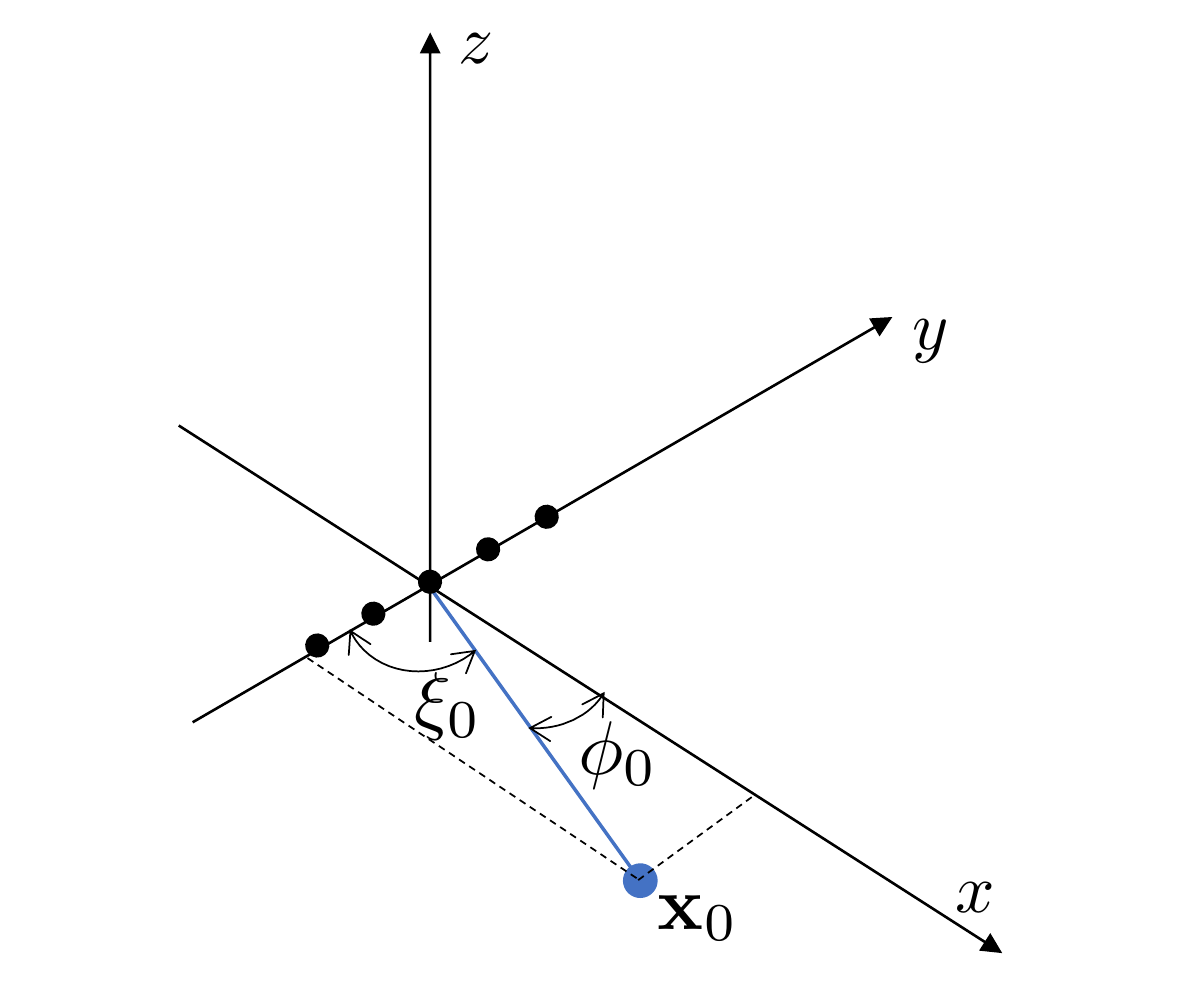}\label{subfig:ULA_geometry}}\\
    \subfloat[][SISO SAR]{\includegraphics[width=0.7\columnwidth]{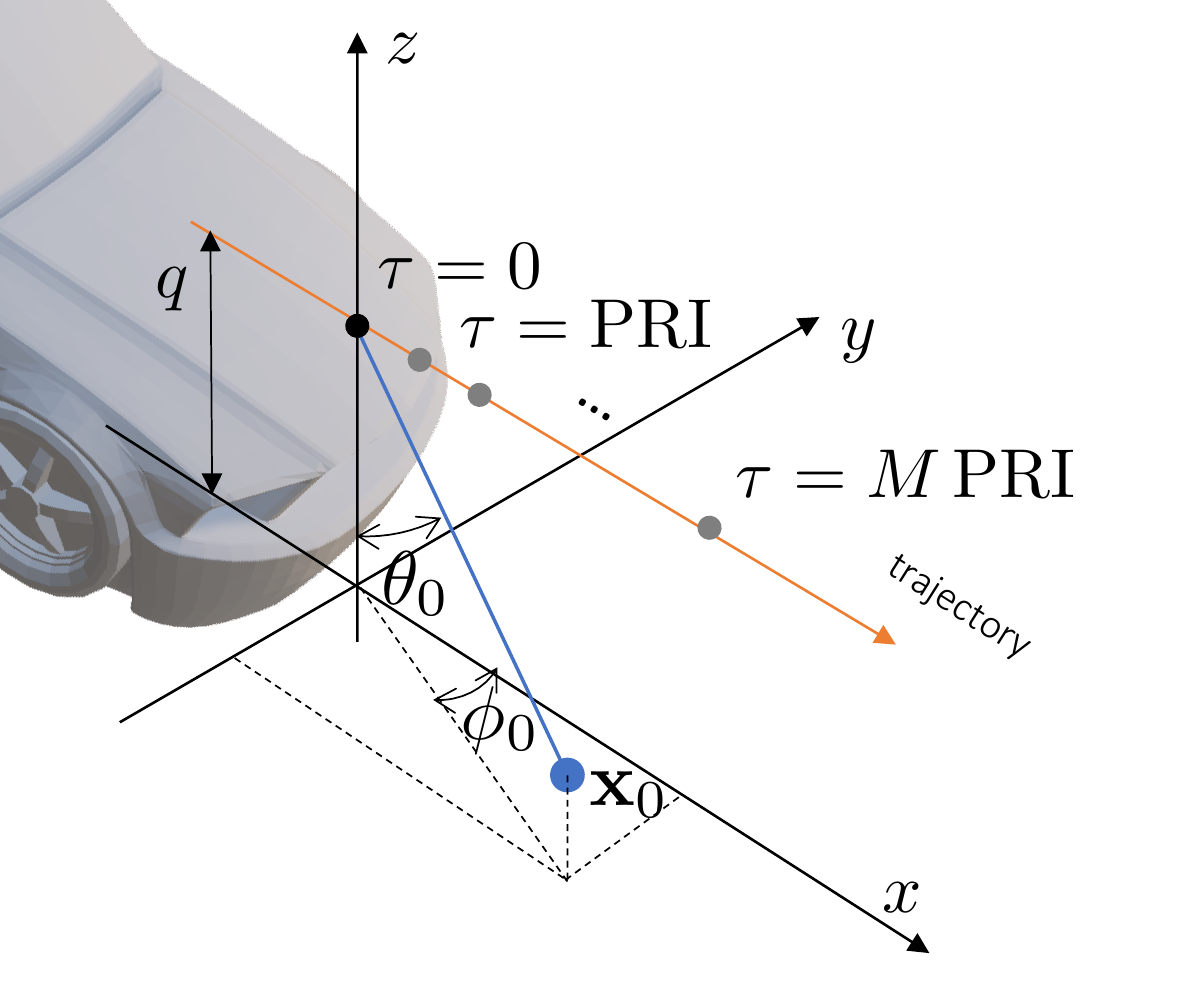}\label{subfig:SISO_SAR}}\\
    \subfloat[][MIMO SAR ]{\includegraphics[width=0.7\columnwidth]{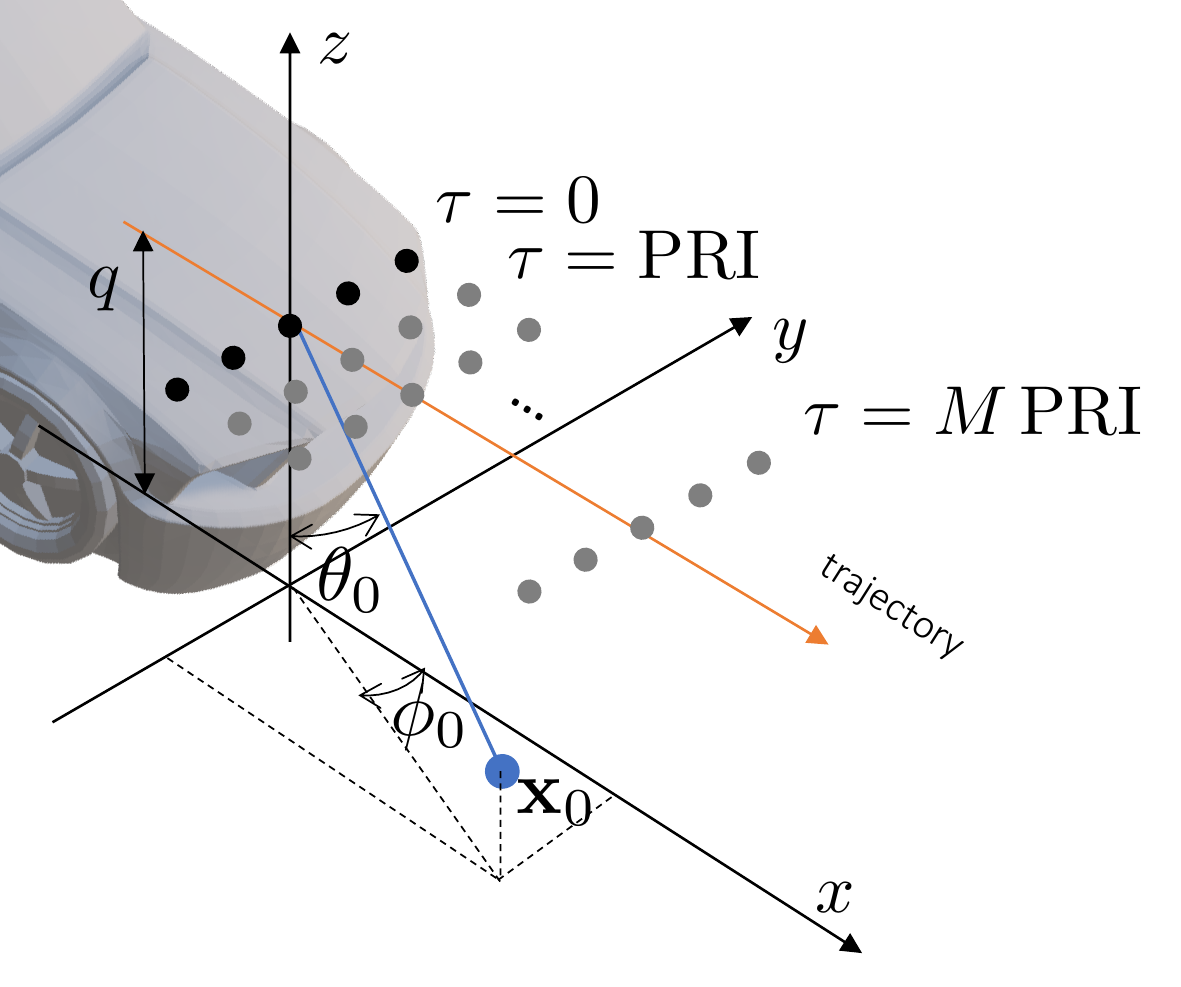}\label{subfig:MIMO_SAR}}
    \caption{Geometry of the radar acquisition: (a) MIMO radar; (b) SISO SAR, at height $q$ from ground, consisting of a single-antenna moving platform, transmitting a pulse for each PRI; (c) MIMO SAR with antenna array orthogonal to motion (forward looking configuration). Notice that other deployments of the radar are possible (for arbitrary squint angles) }
    \label{fig:geometry}
\end{figure}

In the case of a real array, $N$ physical antennas transmit a signal and receive the echo, while in the virtual array scenario there are $N_{tx}$ transmitting antenna and $N_{rx}$ receiving ones. Each possible pair of TX-RX form a virtual radar channel leading to an equivalent array of $N = N_{tx} \times N_{rx}$ virtual elements. From now on, we will generically refer to antenna arrays, without explicitly specify if real or virtual.
Figure \ref{subfig:ULA_geometry} shows a Uniform Linear Array (ULA) displaced along $y$, with an inter-antenna spacing of $\Delta y$. The direction of motion is identified by $x$ axis, i.e., the MIMO radar is in a forward looking configuration, to be consistent with the experimental settings of Section \ref{sec:results}.

Let us therefore consider a 2D scenario. According to \eqref{eq:range_compressed_FMCW}, the Rx signal at the $n$-th antenna from the target in $\mathbf{x}_0$ is:
\begin{equation}
\label{eq:range_compressed}
\begin{split}
        s_{rc}(r,n;\mathbf{x}_0) = T_p\, \mathrm{sinc}&\left[\frac{r-r(n;\mathbf{x}_0)}{\rho_r}\right]\times\\
       & \times \exp\left\{-j\frac{4 \pi}{\lambda}r(n;\mathbf{x}_0)\right\}
\end{split}
\end{equation}
where $r(n;\mathbf{x}_0)$ is the distance from the target in $\mathbf{x}_0$ to the $n$-th antenna. Assuming a plane wave impinging the antenna array, it is:
\begin{equation}
\label{eq:distances}
    r(n;\mathbf{x}_0) \approx r_0 - n \Delta y \sin\phi_0 
\end{equation}
where $r_0 = r(0;\mathbf{x}_0)$ is the distance between the target and the center of the array and $\phi_0 = \tan^{-1}(y_0/x_0)$ is the observation angle (angular position of the target in the FOV).
Combining \eqref{eq:range_compressed} with \eqref{eq:distances}, we obtain:
\begin{equation} \label{eq:range_compressed_ula}
\begin{split}
    s_{rc}(r, n; \mathbf{x}_0&)  = T_p\, \mathrm{sinc}\left[\frac{r-r_0}{\rho_r}\right] \times \\
    & \times \exp\left\{-j\frac{4 \pi}{\lambda}r_0\right\} \exp\left\{j\frac{4 \pi}{\lambda} n \Delta y\sin\phi_0\right\},
\end{split}
\end{equation}
i.e., a truncated spatial sinusoid across the array of frequency $f_x = (2/\lambda)\sin\phi_0$. Therefore, the azimuth compression (or DOA estimation) is again a frequency estimation problem. As the maximum spatial frequency on the array is $f_{x,\mathrm{max}} = 2/\lambda$ (for $\phi_0 = \pm 90$ deg), it follows that the minimum inter-antenna spacing shall be constrained to $\Delta y_{\mathrm{min}} = \lambda/4$. The angular resolution of the array is:
\begin{equation}
\label{eq:angular_resolution_real}
    \rho_\phi = \frac{\lambda}{2 N\Delta y\,\cos\phi} \quad [\mathrm{rad}]
\end{equation}
For instance, a $N=5$ element array displaced by $\Delta y = \lambda/4$ obtains a maximum resolution at boresight of just $23$ deg. In Figure \ref{fig:mimo_image}, we show an example from real data of a MIMO image acquired by an 8 channel array: the angular resolution is roughly $15$ deg.
To improve angular resolution, it is possible to use larger arrays at increased costs and system's complexity or, as here, exploiting the vehicle motion to synthesize a longer array leading to a much finer angular resolution. %\textbf{Qui va messa l'espressione dell'immagine MIMO, che serve poi dopo per il MIMO SAR}

\subsection{SAR Processing}

%The final goal of the automotive radar imaging is to resolve the 2D/3D positions of the targets with a finer resolution compared to standard MIMO radars.
The core of SAR is to jointly process several radar pulses gathered by a radar mounted on a moving platform. For the sake of simplicity, let us consider a single antenna moving on a platform and a target in $\mathbf{x}_0 = [x_0,y_0,z_0]^\mathrm{T}$ (3D scene). The geometry of the problem is depicted in Figure \ref{subfig:SISO_SAR}.
%
%\begin{figure}[!t]
 %   \centering
 %   \includegraphics[width=\columnwidth]{Images/geometry.pdf}
%    \caption{Geometry of the SAR acquisition. The antenna is mounted on a moving platform at an height $q$ from ground, and it is transmitting a pulse every PRI seconds. For a MIMO SAR we have more than one APC for each slow time.}
 %   \label{fig:geometry}
%\end{figure}
%
The RC signal can be written by substituting in \eqref{eq:range_compressed} the antenna index $n$ with the \textit{slow-time} $\tau$:
\begin{equation}
\label{eq:SAR_RC}
\begin{split}
 s_{rc}(r, \tau ; \mathbf{x}_0) = T_p\, &\mathrm{sinc}\left[\frac{r-r(\tau, \mathbf{x}_0)}{\rho_r}\right]\times\\ &\times\exp\left\{-j\frac{4 \pi}{\lambda}r(\tau, \mathbf{x}_0)\right\}.
\end{split}
\end{equation}
There are several algorithms that can be used for the so-called \textit{focusing}. The most adequate for non linear trajectories is the Time Domain Back Projection (TDBP) \cite{cumming_digital_2005, yu_signal_2020}. The TDBP integral for a generic pixel in the scene $\mathbf{x} = [x,y,z]^{\mathrm{T}}$ can be written as:
\begin{equation}
\label{eq:TDBP_sum}
    I(\mathbf{x}) = \int\limits_{\tau \in T} s_{rc}(r(\tau,\mathbf{x}), \tau; \mathbf{x}_0) \exp\left\{j\frac{4\pi}{\lambda}r(\tau,\mathbf{x})\right\} \mathrm{d}\tau
\end{equation}
where $I(\mathbf{x})$ is the final SAR image, $T$ is the considered synthetic aperture time and $r(\tau;\mathbf{x}) = \|\mathbf{x}-\mathbf{p}(\tau)\|$ is the time-varying antenna-to-pixel distance at a given time $\tau$, function of the position of the Antenna Phase Center (APC) represented by the vector $\mathbf{p}(\tau) = [p_x(\tau),p_y(\tau),p_z(\tau)]^{\mathrm{T}}$.

\begin{figure}[!t]
\centering
\includegraphics[width=0.9\linewidth]{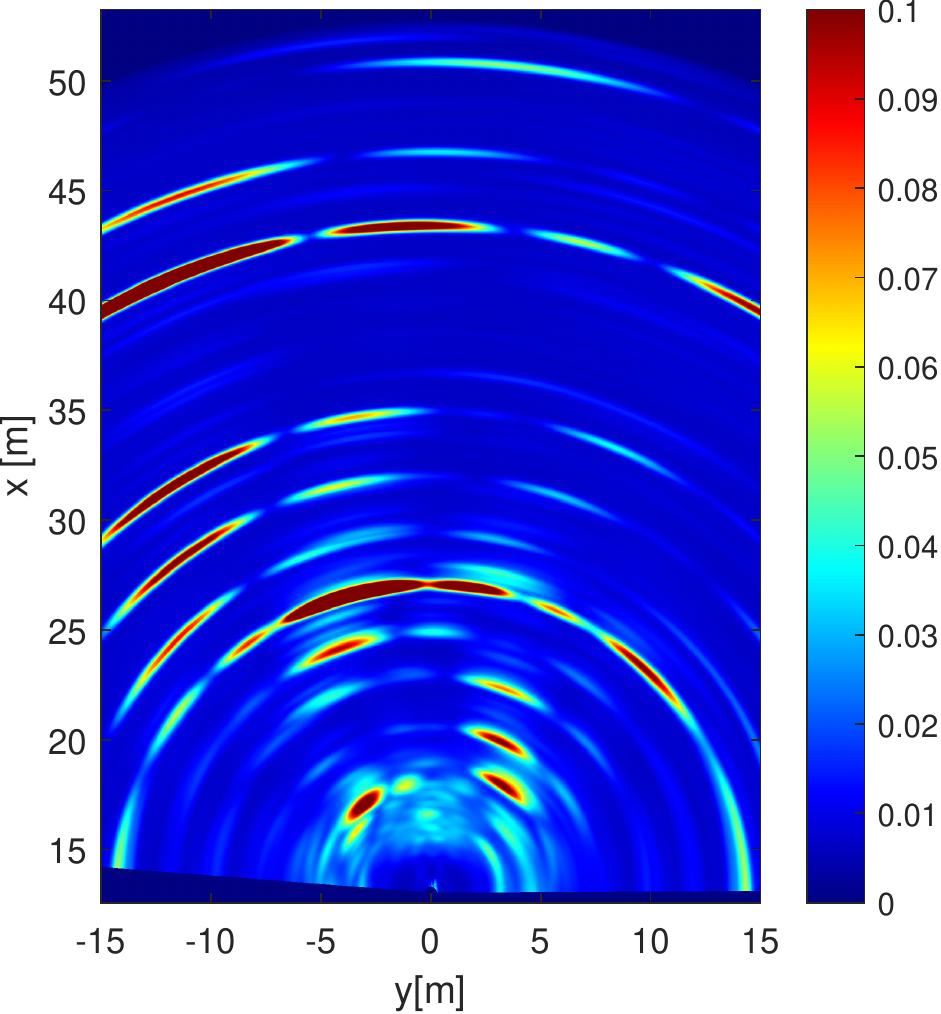}
\caption{MIMO image acquired by an 8 channel array. With 8 elements spaced by $\lambda/4$ the angular resolution is roughly $15$ deg.}
\label{fig:mimo_image}
\end{figure}

The whole TDBP algorithm is divided in three steps: \textit{(i)} the RC data for a single pulse is evaluated at position $r(\tau;\mathbf{x})$ \textit{(ii)} the interpolated data is phase rotated by $\exp\left\{j(4\pi/\lambda) r(\tau;\mathbf{x})\right\}$ compensating for the two-way path phase \textit{(iii)} the procedure is repeated for every pulse into the synthetic aperture and the results are coherently summed.
A very simple interpretation of the TDBP algorithm can be given if we assume a 2D geometry and a rectilinear trajectory of the platform with constant velocity. Let us therefore assume the platform traveling at ground level (i.e, $q = 0$ or $\theta = 90$ deg for all the pixels in the FoV) along the $x$ axis with a velocity $\mathbf{v} = [v_x,0,0]$. If the target is located at a generic $\mathbf{x}_0 = [x_0,y_0,0]^\mathrm{T}$, we have:
\begin{equation}
\label{eq:range_approx}
    r(\tau;\mathbf{x}) \approx r_0 + \xi v_x \tau
\end{equation}
where $\xi = \pi/2-\phi$, the TDBP integral \eqref{eq:TDBP_sum} reduces to:
\begin{equation}
\label{simplified_TDBP}
\begin{split}
        I(\mathbf{x}) &\approx \int\limits_{\tau \in T} C  \exp\left\{-j\frac{4 \pi}{\lambda}(\xi-\xi_0)v_y\tau\right\} \mathrm{d}\tau \approx \\
        &  \approx T\,\mathrm{sinc}\left[\frac{2 v_x T}{\lambda}(\xi-\xi_0)\right]
\end{split}
\end{equation}
where we assume a constant illumination of the target along the whole aperture time, i.e., $|s_{rc}(r(\tau,\mathbf{x}),\tau,\mathbf{x}_0)|=C$. The expression of \eqref{simplified_TDBP} is the Fourier transform of a truncated complex sinusoid with frequency $f_d = 2 v_x \xi_0/\lambda$, therefore the result will be a cardinal sine function in the Doppler frequency domain centered in $f_d$.

\begin{figure}[!t]
\centering
\includegraphics[width=0.9\linewidth]{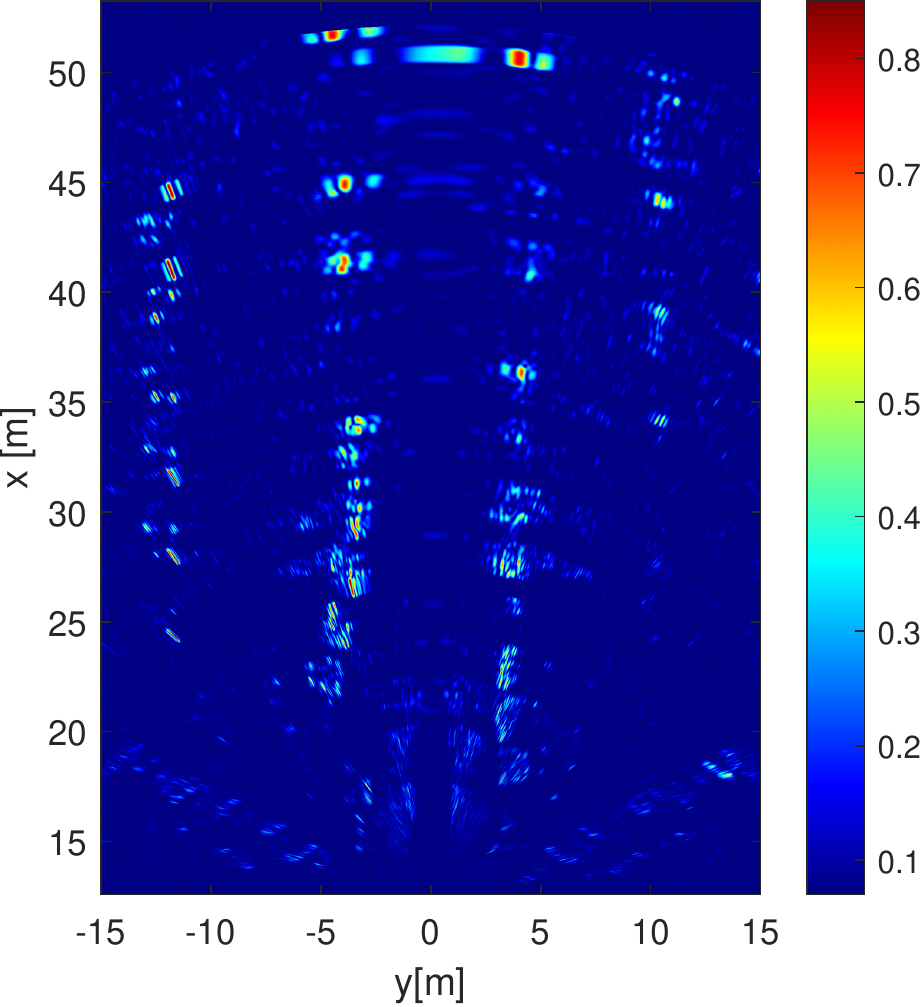}
\caption{SAR image focussed using an aperture length of $T =150$ ms. The angular resolution is greatly improved with respect to the standard MIMO image. The amplitudes are normalized and in linear scale.}
\label{fig:sar_image}
\end{figure}

Up to now we derived the SAR processing for a Single-Input-Single-Output (SISO) architecture. In MIMO systems, instead, we have multiple APCs (real or virtual) travelling on the same platform, as exemplified in Figure \ref{subfig:MIMO_SAR} for a forward looking configuration (other configurations, such as side-looking, are also possible). To have a MIMO SAR, we first observe that at each slow time $\tau$ we can form a low resolution MIMO image by simple spectral analysis (see Section \ref{sec:ULA}) or by TDBP leading to:
\begin{equation}
\label{eq:TDBP_MIMO}
    I_m(\mathbf{x},\tau) \hspace{-0.1cm}=\hspace{-0.1cm} \sum_{n=1}^N s_{rc}[r(n,\tau ; \mathbf{x}),n,\tau;\mathbf{x}_0]\mathrm{exp}\left\{j\frac{4\pi}{\lambda}r(n,\tau ; \mathbf{x})\right\},
\end{equation}
where $I_m(\mathbf{x},\tau)$ is the low resolution image obtained by the focusing of the $N$ signals received at time instant $\tau$. The final SAR image is then obtained by coherently summing all the complex-valued low-resolution MIMO images along the synthetic aperture:
\begin{equation}
\label{eq:sum_MIMO}
    I(\mathbf{x}) = \sum_{\tau \in T}I_{m}(\mathbf{x},\tau ).
\end{equation}
The angular resolution of SAR systems improves significantly compared to conventional MIMO radars, where the effective aperture $N \Delta y$ is substituted by the synthetic aperture length $A_s = v_x\tau$:
\begin{equation}
    \rho_\phi^{\text{SAR}} = \frac{\lambda}{2A_s \sin\phi} \quad [\mathrm{rad}]
\end{equation}

The angular resolution is then converted into spatial resolution in the direction of motion with a linear relationship $\rho_x \approx r \rho_\phi^{\text{sar}}$. Notice that the maximum spatial resolution of a MIMO radar is for $\phi=0$ deg, while for a SAR corresponds to $\phi=90$ deg (orthogonal to the synthetic aperture). For a system operating at $77$ GHz munted on a vehicle traveling at $14$ m/s ($54$ km/h), we obtain a resolution of $0.2$ deg at the synthetic aperture boresight by exploiting $50$ cm of aperture. The cross-range spatial resolution is $11$ cm at $r=30$ m of distance. Figure \ref{fig:sar_image} shows the same scene of Figure \ref{fig:mimo_image}, but this time exploiting the car's motion to form a synthetic aperture of $T=150$ ms ($A_s \approx 90$ cm). Notice that Figure \ref{fig:sar_image} is a normalized SAR image (i.e., the amplitude value is between 0 and 1). The angular resolution is greatly improved allowing for better recognition and localization of the targets. The usage of such a synthetic aperture is equivalent to an equivalent array of $\approx 930$ real/virtual channels.

\section{Motion error analysis}
\label{sec:motion_error}

\begin{figure}[!t]
    \centering
    \includegraphics[width=0.9\columnwidth]{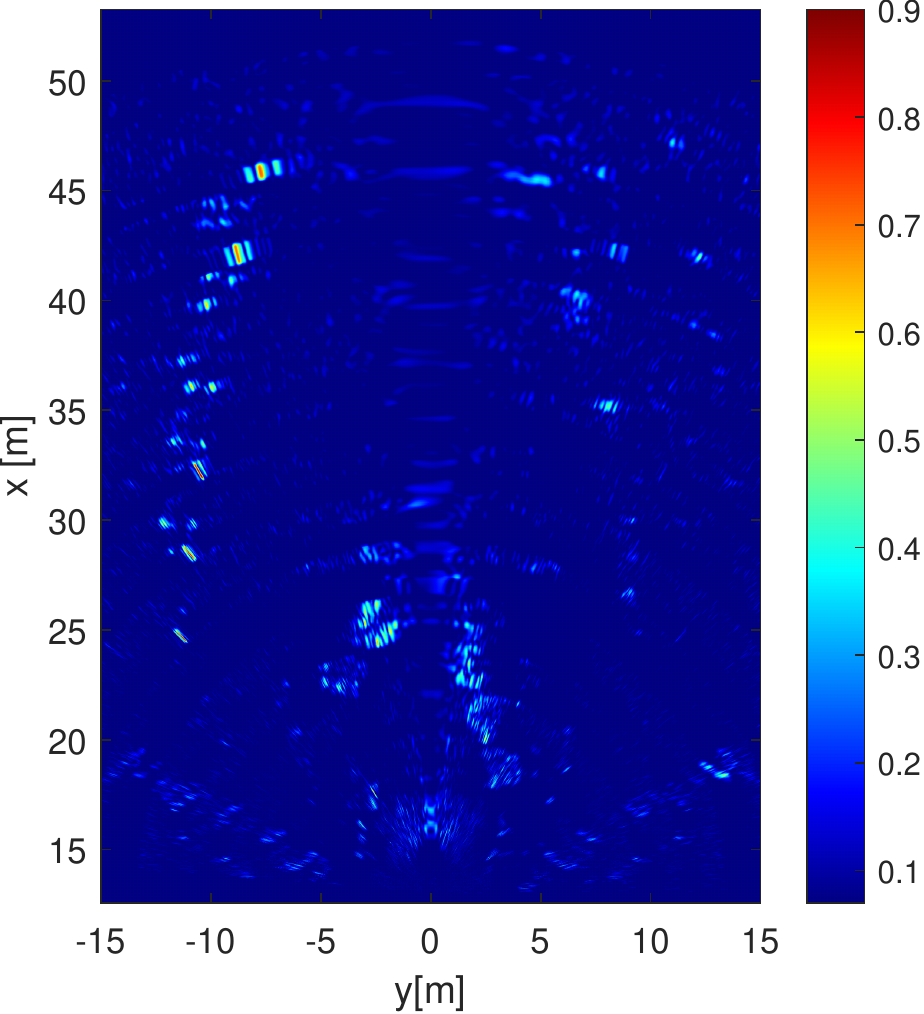}
    \caption{The same scene of Figure \ref{fig:sar_image}, but now corrupted by a severe trajectory error along $x$. The image collapses inwards. ($\Delta v_x = 35$ cm/s). The amplitudes are normalized and in linear scale.}
    \label{fig:error_x}
\end{figure}

We recall that each SAR processor, described by \eqref{eq:TDBP_sum}, requires the knowledge of the APC positions at each slow time $\tau$ for the computation of the range values $r(\tau; \mathbf{x})$. In principle, the vehicle trajectory must be known with an accuracy within the wavelength (millimeters for typical automotive radars). In practice, it is sufficient to track the \textit{relative} APC motion along a synthetic aperture, i.e., to know the position \textit{displacement}. In this section, we focus on velocity errors, as stationary position errors do not affect the quality of the SAR images, while linear position errors due to velocity errors lead to image distortion. We first analytically derive the effect of a velocity error on the focused SAR image $I(\mathbf{x})$, then we set the theoretical requirement on velocity estimation accuracy, discussing the implications for typical automotive SAR systems and justifying the usage of both navigation and radar data to properly perform the residual motion estimation and compensation.

\begin{figure}[!t]
    \centering
    \includegraphics[width=0.9\columnwidth]{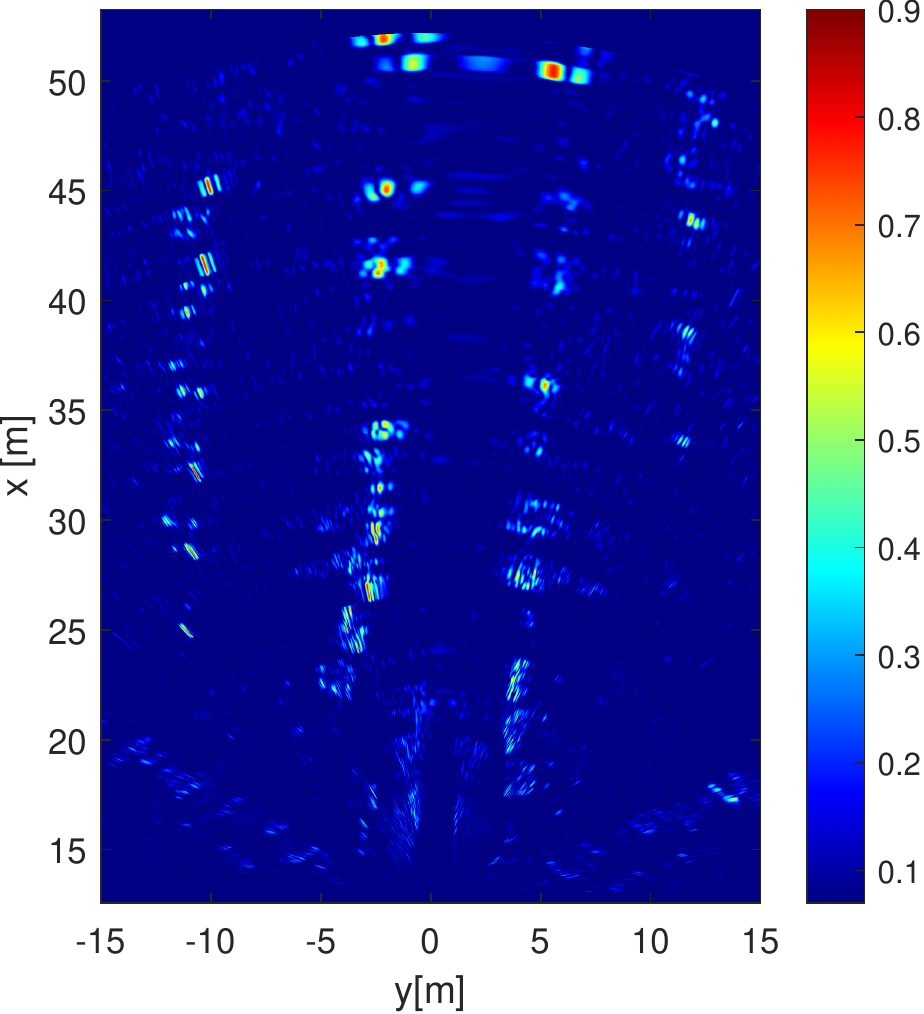}
    \caption{The same scene of Figure \ref{fig:sar_image}, but now corrupted by a severe trajectory error along $y$. The image is rotated ($\Delta v_y = 35$ cm/s). The amplitudes are normalized and in linear scale.}
    \label{fig:error_y}
\end{figure}

In the following, we model the navigation output as a noisy biased estimate of the true vehicle's position $\mathbf{p}(\tau)$ and velocity $\mathbf{v}(\tau)=[v_x(\tau),v_y(\tau),v_z(\tau)]^\mathrm{T}$. Let us define the vehicle's instantaneous velocity provided by the navigation system as:
\begin{equation}
    \mathbf{v}_{nav}(\tau) = \mathbf{v}(\tau) + \Delta\mathbf{v}(\tau),
\end{equation}
where $\Delta\mathbf{v}(\tau) = [\Delta v_x(\tau), \Delta v_y(\tau), \Delta v_z(\tau)]^\mathrm{T}$ is the velocity error. From hereafter, we assume that the vehicle is travelling at constant speed $\mathbf{v}$ within a single synthetic aperture, namely we neglect the acceleration. This assumption is reasonable in typical urban scenarios, where the velocity is limited, the dynamics of the vehicle is moderate, and $A_s$ ranges from few to tens of centimeters. For the same reason, we also consider a constant velocity estimation error $\Delta \mathbf{v}$. Therefore, the synthetic aperture can be approximated as 
\begin{equation}
    A_s = \int\limits_{\tau \in T}\|\mathbf{v}(\tau)\|
 \,\mathrm{d}\tau \approx \|\mathbf{v}\|\,T.
\end{equation}
From \eqref{eq:range_approx}, we observe that a constant velocity error maps to a linear range over time, and therefore to a linear phase. If the trajectory is perfectly known ($\Delta\mathbf{v}=\mathbf{0}$), the complex exponential in the TDBP integral of \eqref{eq:TDBP_sum} will perfectly compensate the phase term of \eqref{eq:SAR_RC} on the pixel $\mathbf{x}_0$ where the target is located, leading to a constructive sum and to a well focused image.
Conversely, if the estimate of the trajectory contains an error $\Delta\mathbf{v}\neq\mathbf{0}$, the phase term in \eqref{eq:SAR_RC} will not be perfectly compensated in \eqref{eq:TDBP_sum}, thus leading to a destructive sum and to a \textit{defocused} image. Moreover, there is an angular displacement of the target in the SAR image leading to a wrong localization.

To gain insight on the role of velocity errors in automotive SAR imaging, consider a 2D geometry, with a vehicle travelling along $x$ at ground level ($q=0$) at velocity $v_x$ and a target placed in $\mathbf{x}_0=[0,r_0,0]^\mathrm{T}$. For a velocity error in the direction orthogonal to the motion ($\Delta v_y \neq 0$), the range expression \eqref{eq:range_approx} for small angles $\xi$ becomes:
\begin{equation}
    r(\tau;\mathbf{x}) \approx r_0 + \xi v_x \tau + \Delta v_y \tau
\end{equation}
and the TDBP \eqref{simplified_TDBP} can be then rewritten as:
\begin{equation}
\label{eq:simplified_TDBP_error}
     I(\mathbf{x}) \approx \hspace{-0.25cm}\int\limits_{\tau \in T} \hspace{-0.15cm} C  \exp\left\{-j\frac{4 \pi}{\lambda}\left(\xi-\xi_0-\frac{\Delta v_y}{v_x}\right)v_x\tau\right\} \mathrm{d}\tau,
\end{equation}
that is again a Fourier integral resulting in a sinc function, but now centered in $\xi_0 - (\Delta v_y/v_x)$.
The error is in the angular localization of the target, $\Delta \xi = \Delta v_y/v_x$, converts into a position error:
\begin{equation}
    \Delta x = r_0\Delta \xi, 
\end{equation}
hindering the precise target localization especially for medium/long ranges. For instance, in Figure \ref{fig:error_x} we image the same scenario of Figure \ref{fig:sar_image}, but corrupted by a strong velocity error along the direction of motion $\Delta v_x = 35$ cm/s. The SAR image seems to collapse inward, with defocused and mis-localized targets. A similar observation can be made for Figure \ref{fig:error_y}, where the velocity error is in the direction orthogonal to the nominal motion ($\Delta v_y = 35$ cm/s). The scene is rotated as predicted by \eqref{eq:simplified_TDBP_error}.
\begin{figure}[!t]
\centering
\includegraphics[width=0.6\linewidth]{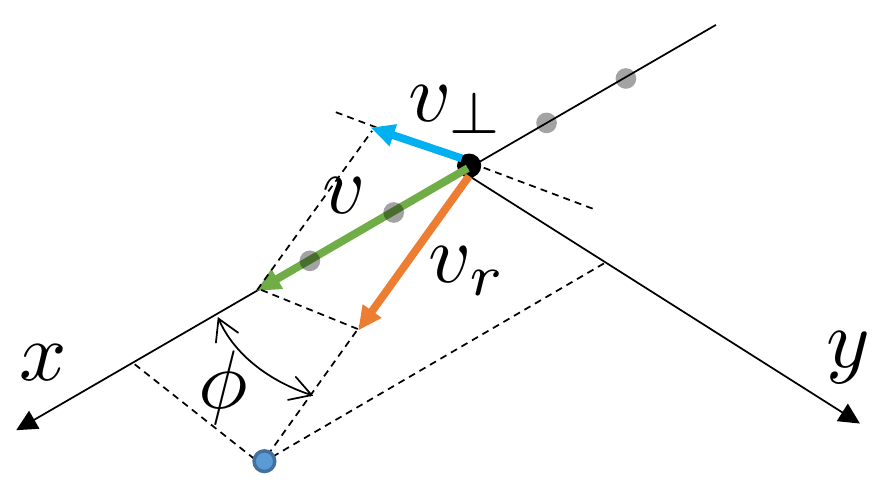}
\caption{Nominal velocity ($\mathbf{v}$), radial velocity $v_r$ orthogonal velocity $v_\perp$. For the sake of simplicity we depicted the 2D geometry.}
\label{fig:radial_orthogonal_velocities}
\end{figure}
The analysis can be generalized to the 3D domain with target position $\mathbf{x} = [x,y,z]^{\mathrm{T}}$ (at range $r=\sqrt{x^2+y^2+z^2}$ from the center of the synthetic aperture) and 3D velocity of the vehicle $\mathbf{v} = [v_x,v_y,v_z]^{\mathrm{T}}$. The phase of the received signal in \eqref{eq:SAR_RC} can be linearized as:
%It is possible to express (excluding a constant term) the phase of the received signal at time $\tau$ by expanding in Taylor series the phase equation:
%
\begin{equation}
\label{eq:linear_phase}
    \begin{split}
        \psi(\mathbf{x}, \tau) &\approx \frac{4\pi}{    \lambda}r-\left(\mathbf{k}(\mathbf{x})^{\mathrm{T}}\mathbf{v}\right) \tau
    \end{split}
\end{equation}
where 
\begin{equation}
\begin{split}
      \mathbf{k}(\mathbf{x}) & = \frac{4\pi}{\lambda}[\sin\theta\cos\phi, \, \sin\theta\sin\phi,\, \cos\theta]^\mathrm{T}. 
\end{split}
\end{equation}

%
%
%\begin{equation}
%\begin{split}
%            \mathbf{k}(\mathbf{x}) &= \frac{4\pi}{\lambda}\left[x/r \quad y/r \quad z/r\right]^\mathrm{T}\\
%            &=\frac{4\pi}{\lambda}[\mathrm{sin}(\theta)\mathrm{cos}(\phi) \quad \mathrm{sin}(\theta)\mathrm{sin}(\phi) \quad \mathrm{cos}(\theta)]^\mathrm{T}\\
%    & = [k_x \quad k_y \quad k_z]^\mathrm{T}
%\end{split}
%\end{equation}
%
%and
%
%\begin{equation}
%    \mathbf{v} = [v_x \quad v_y \quad v_z]^\mathrm{T}
%\end{equation}
%
%
\begin{figure}[!t]
    \centering
    \includegraphics[width=0.9\columnwidth]{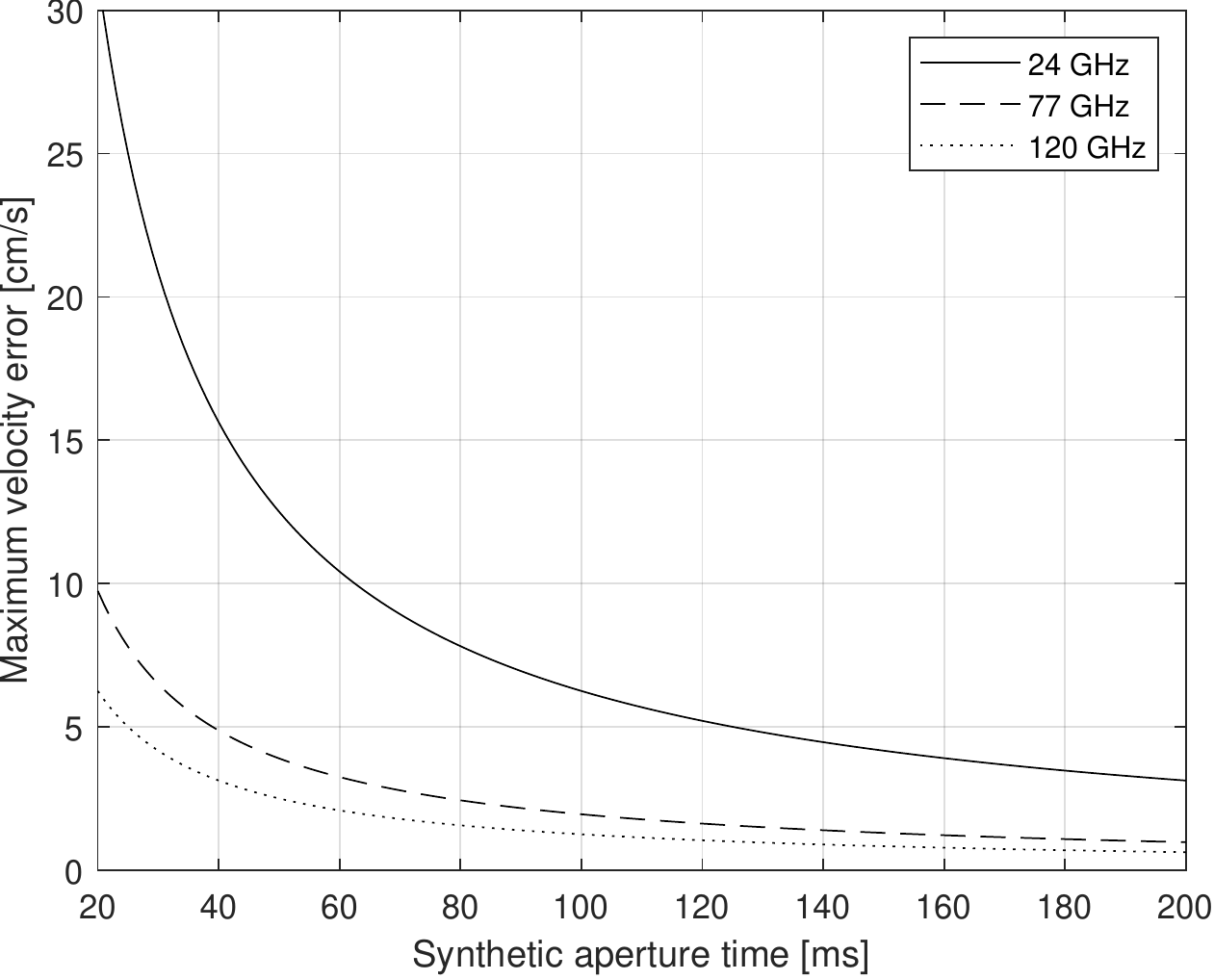}
    \caption{The maximum tolerable velocity error in function of the total integration (aperture) time. The higher the integration time, the more stringent are the constraints on the knowledge of the car's velocity. The tolerances are evaluated for different operational frequencies of the radar.}
    \label{fig:maximum_error}
\end{figure}
An error in the velocity is therefore transferred into a phase error
\begin{equation}
\label{eq:residual_phase}
\begin{split}
        \Delta\psi^v(\mathbf{x}, \tau) &= \frac{\partial \psi(\mathbf{x}, \tau)}{\partial \mathbf{v}}\Delta \mathbf{v} = \left(\mathbf{k}(\mathbf{x})^\mathrm{T} \Delta \mathbf{v}\right) \tau.
\end{split}
\end{equation}
%
%showing that, for localization purposes, SAR imaging is therefore maximally sensitive to velocity errors in the target's observation direction (defined by $\mathbf{k}(\mathbf{x})$), while any velocity error orthogonal to $\mathbf{k}(\mathbf{x})$ leads to an image defocusing. It is important to underline that trajectory errors are not the only sources of a linear residual phase: for instance, an error in the focusing height $h$ leads again to a linear residual phase. In general, however, the velocity error from the navigation system dominates the phase error. 
When $\Delta \mathbf{v} = \mathbf{0}$, the residual phase $\Delta\psi^v(\mathbf{x}, \tau)$ over a target at $\mathbf{x}$ is zero; conversely, when $\Delta \mathbf{v} \neq \mathbf{0}$, the phase shows a linear behavior with $\tau$, representing a residual Doppler frequency. It is also interesting to notice that the phase error is higher in those areas of the FOV pointed by vector $\Delta \mathbf{v}$. If the dominant contribution of the velocity error is in the direction of motion $x$, for instance, the area of the final image that will be more corrupted by the velocity error is the one around $x$.

It is now useful to assess the maximum tolerable velocity error. From \eqref{eq:simplified_TDBP_error}, a velocity error results in a positioning error in the final SAR image. The maximum tolerable velocity error is defined as the one that induces a localization error within the angular resolution. Eq. \eqref{eq:simplified_TDBP_error} can be extended as:
\begin{equation}
    \label{eq:max_radial_velocity}
    \frac{\Delta v_r^{\text{max}}}{v_{\perp}} = \frac{\lambda}{2A_s^\perp}
\end{equation}
where $\Delta v_r^{\text{max}}$ is the maximum tolerable \textit{radial} velocity error in an arbitrary direction defined by $(\theta,\phi)$ (line of sight), $v_{\perp}$ and $A_s^\perp$ are, respectively, the nominal vehicle's velocity and the component of the synthetic aperture orthogonal to the line of sight. In Figure \ref{fig:radial_orthogonal_velocities} the geometry of acquisition is depicted with the quantities just mentioned highlighted. We can also express the absolute tolerable velocity by recognizing that $A_s^\perp = v_{\perp}T$, where $T$ is the integration time, thus:
\begin{equation}\label{eq:radial_velocity_max}
    \Delta v_r^{\text{max}} = \frac{\lambda}{2T}
\end{equation}
Figure \ref{fig:maximum_error} depicts the maximum tolerable radial velocity error as function of the total integration time and for different operational wavelengths. For a long integration time (i.e, thus high azimuth resolution) and shorter wavelengths, the requirements on the accuracy become very strict, in the order of $1$ cm/s. For instance, for a car moving at $15$ m/s, an angular resolution of $\rho_\phi^{\text{sar}}=0.2$ deg at $\phi = 90$ deg (allowing $1$ m of cross-range resolution at $30$ m at $77$ GHz) implies $T\approx 40$ ms, therefore the velocity error shall be within $v_r^{\text{max}}=5$ cm/s. Notice that, reducing the frequency of operation, e.g., to $24$ GHz, does not relax the requirement: an angular resolution of $\rho_\phi^{\text{sar}}=0.2$ deg requires an aperture time $T\approx 125$ ms, leading again to $\Delta v_r^{\text{max}} \approx 5$ cm/s.
Automotive-legacy navigation systems can provide an average velocity error ranging from $5$ cm/s down to $2-3$ cm/s for expensive commercial Real-Time Kinematic (RTK) setups, possibly integrating GNSS, inertial sensors and magnetometers~\cite{InertialSense2020}. However, these systems heavily rely on GNSS signals, that may be absent or inaccurate in some scenarios (urban canyons with strong multipath or tunnels). Moreover, car navigation systems must deal with unpredictable driver's maneuverings, leading to a velocity error that could be as high as $10-20$ cm/s \cite{novatel_compact_2018}. Although the fusion of inexpensive heterogeneous in-car sensors data was demonstrated to provide accurate imaging in moderate dynamics~\cite{tagliaferri_navigation-aided_2021}, a reliable SAR imaging for autonomous driving applications calls for the integration of navigation \textit{and} radar data.

%A trajectory error, however, is not the only one that manifests as a linear residual phase. Another common cause is an error in the focusing height.

%In \eqref{eq:residual_phase}, we evaluate the effect of a velocity error $\Delta \mathbf{v}$ in the phase of the backprojected signal. The result is a linear residual phase with $\tau$. In this section we show that an error in the focusing height can also be responsible for a linear residual phase term.\\

%%%%%%%%%%%%%%% MOCO %%%%%%%%%%%%%%%%%%%%%%%%%%
\section{Motion error estimation and compensation}
\label{sec:moco} 
\begin{figure*}[!t]
\centering
\includegraphics[width=\linewidth]{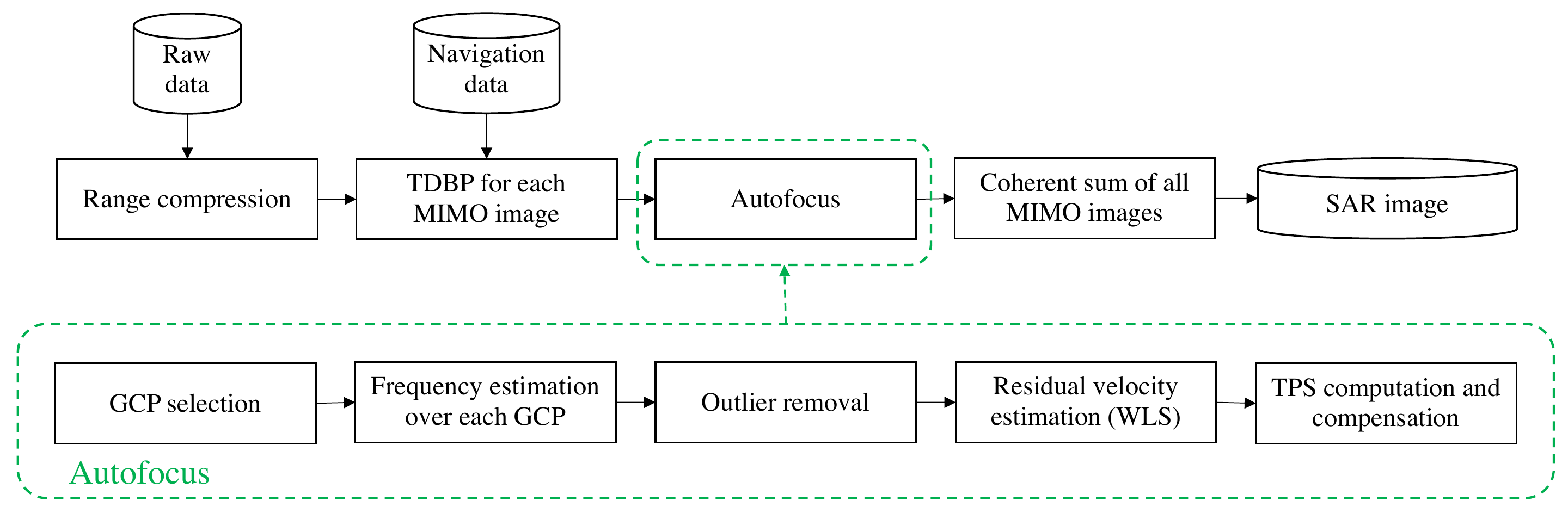}
\caption{Block diagram of the SAR focusing pipeline. The autofocusing procedure is a key part of the workflow.}
\label{fig:block_diagram}
\end{figure*}

This section outlines the proposed residual motion estimation and compensation technique to estimate and compensate trajectory errors starting from a set of low resolution MIMO images $\{I_m(\mathbf{x},\tau)\}_{\tau\in T}$. After the estimation of the motion error, each MIMO image is first corrected by a phase term referred to as Trajectory Phase Screen (TPS), then a well-focused SAR image is obtained by \eqref{eq:sum_MIMO}. Figure \ref{fig:block_diagram} shows the complete SAR data processing and autofocus workflow. In the following, we detail each portion of the block diagram and we outline some guidelines for a practical implementation of the algorithm.
\subsection{From raw data to a stack of low-resolution images}
The vehicle is equipped with a MIMO FMCW array able to generate an equivalent array of $N$ elements (either virtual or real). At each slow time, each one of the $N$ elements of the array receives a radar echo. Each received signal is first RC and then subject to the TDBP for the generation of the $M$ low resolution images in the duration of the synthetic aperture $T$: each image $I_m(\mathbf{x},\tau)$ is formed by back-projecting in the FOV the $N$ signals received at each APC of the array at the time instant $\tau$. Notice that the combination of PRI and number of MIMO channels of the radar must ensure to have unambiguous low resolution images.
If the TDBP is performed over a common grid (FOV) for all the slow time instants, all the images are already compensated for range migration. Conversely, if the focusing of the MIMO image is done using a simple Fourier transform as discussed in Section \ref{sec:ULA}, the images must be then co-registered to a common grid through an interpolation step \cite{cumming_digital_2005}.\\
The former approach has been used in this work.

\subsection{Autofocus}
The core of the procedure, i.e., the autofocus routine, takes as input the set of $M$ low resolution images $\{I_m(\tau,\mathbf{x})\}_{\tau\in T}$. As discussed in Section \ref{sec:motion_error}, the presence of a constant error in the estimated velocity $\Delta \mathbf{v}$ by navigation leads to a linear residual range (or phase) after TDBP, leading to a distorted image. The autofocus procedure starts from the linear residual phase $\Delta \psi^v(\mathbf{x}, \tau)$ in \eqref{eq:residual_phase}, representing a complex sinusoid of angular frequency:
\begin{equation}
\label{eq:frequency_residual}
   \Delta \omega(\mathbf{x}) =  \mathbf{k}(\mathbf{x})^\mathrm{T} \Delta\mathbf{v},
\end{equation}
that provides a single equation for three unknowns (the three components of the velocity error).
Exploiting the low-resolution images at each slow time, it is sufficient to detect few stable GCPs in the scene to have an overdetermined linear system of equations. If we consider a total of $P$ GCPs we can write:
\begin{equation}
\label{eq:linear_system}
    \underbrace{\begin{bmatrix}
        \Delta \omega(\mathbf{x}_0)  \\
        \Delta \omega(\mathbf{x}_1)  \\
        \Delta \omega(\mathbf{x}_2)  \\
        \vdots \\
        \Delta \omega(\mathbf{x}_P) 
    \end{bmatrix}}_{\Delta \boldsymbol{\omega}}
    = 
    \underbrace{\begin{bmatrix}
    k_x^0 & k_y^0 & k_z^0\\
    k_x^1 & k_y^1 & k_z^1\\
    k_x^2 & k_y^2 & k_z^2\\
    \vdots & \vdots & \vdots\\
    k_x^P & k_y^P & k_z^P
    \end{bmatrix}}_{\mathbf{K}}
    \underbrace{\begin{bmatrix}
    \Delta v_x \\
    \Delta v_y \\
    \Delta v_z
    \end{bmatrix}}_{\Delta \mathbf{v}}
    +
    \underbrace{\begin{bmatrix}
        n_0  \\
        n_1  \\
        n_2  \\
        \vdots \\
        n_P 
    \end{bmatrix}}_{\mathbf{n}}
\end{equation}

where $\mathbf{n}\sim\mathcal{CN}(\mathbf{0},\sigma^2_n \mathbf{I})$ is the circularly complex Gaussian noise vector of power $\sigma^2_n$ representing the uncorrelated noise on the estimates of the residual frequencies $\Delta \boldsymbol{\omega}$.
The process of selection of stable GCPs can be performed by looking at the amplitude statistics of the scene. In particular, we opt for the computation of the \textit{incoherent} average (i.e., the average of the amplitudes) of all the $M$ low resolution images and take just the brightest targets (i.e., the ones with the higher phase stability). Since the linear system \eqref{eq:linear_system} has only three unknowns, it is typically sufficient to detect $20$ to $50$ GCPs to obtain a reliable estimate. Therefore, the thresholding on the amplitude for the GCP selection can be very stringent.
Another key aspect in the selection of GCP is the accuracy on their localization. An error in the focusing height and/or an error in the estimation of the angular position of the GCPs can prevent a correct estimation of the residual motion error.\\

\begin{figure}[!t]
    \centering
    \includegraphics[width=0.9\columnwidth]{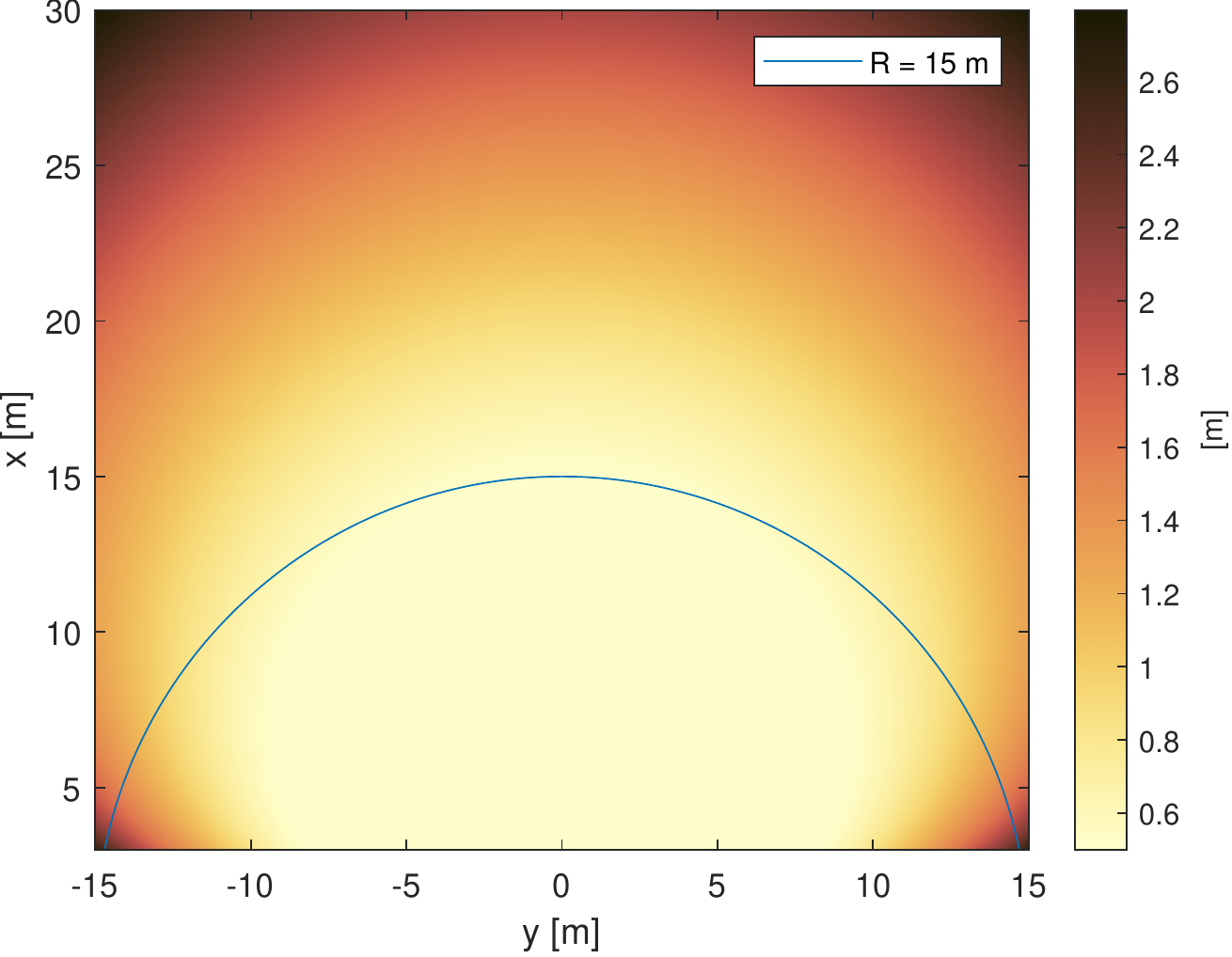}
    \caption{Maximum tolerable focusing height error on GCP. The nominal vehicle velocity is set to 15 m/s along $x$, the nominal radar height over ground is 0.5 m, the maximum radar slant range is 30 m. Notice that in far range, the system becomes tolerable to focusing heights errors.}
    \label{fig:maximum_elevation_error}
\end{figure}

\subsubsection{Error in the focusing height}
In \eqref{eq:TDBP_sum}, the 3D coordinate $\mathbf{x}$ of every pixel of the backprojection grid can be arbitrarily chosen. It is common practice to choose a fixed focusing height, i.e., $\mathbf{x} = [x,y,z=z']^\mathrm{T}$. In a real scenario, however, the height of the target is not known and it is possibly different from $z'$ \cite{duque_precise_2019}. The difference between the true position and the focusing plane is denoted by $\Delta z$. If the target is not truly at that position, another linear residual phase term arises. 
The expression of such residual phase can be easily derived considering that an error on $z$ is equivalent to an error on the elevation angle $\theta$, thus:
\begin{equation}
\label{eq:residual_phase_elevation}
\begin{split}
        \Delta \psi^z(\mathbf{x}, \tau) &= \frac{\partial \psi(\mathbf{x}, \tau)}{\partial \theta}\Delta \theta = \left[\left(\mathbf{k}'_\theta(\mathbf{x})^\mathrm{T} \mathbf{v} \right) \Delta \theta\right]  \tau
\end{split}
\end{equation}
where $\mathbf{k}_\theta'(\mathbf{x})=(4\pi/\lambda)[\cos\theta\cos\phi, \, \cos\theta\sin\phi,\, -\sin\theta]^\mathrm{T}$ is the derivative of $\mathbf{k}(\mathbf{x})$ with respect to $\theta$ and 
\begin{equation}\label{eq:elevation_error}
    \Delta \theta = \frac{\Delta z}{r\sin\theta}.
\end{equation}
Notice that \eqref{eq:residual_phase_elevation} is again linear in slow time: both an error in the velocity and an error in the focusing height manifest as a linear phase in time. 
The residual linear phase over a GCP depends on the nominal velocity of the vehicle $\mathbf{v}$, the angles $(\theta,\phi)$, the range $r$ and the error of focusing height $\Delta z$.
It is now possible to compute the maximum tolerable error in term of $\Delta z$ which will again depend on the GCP position $(r,\theta,\phi)$ and on the vehicle velocity $\mathbf{v}$.
Exploiting the requirement on the radial velocity \eqref{eq:radial_velocity_max}, the maximum tolerable height error is the one that generates a Doppler frequency corresponding to $\Delta v_r^{\text{max}}$. In Figure \ref{fig:maximum_elevation_error}, we show the maximum tolerable height error for $77$ GHz radar with $T=200$ ms integration time, represented for each pixel in the scene. Notice that, according to \eqref{eq:elevation_error}, in far range the system is very robust even to big elevation errors. This is a direct consequence of large $r$ and steep incidence angles ($\theta \approx 90$ deg). The suggestion is then to select GCPs in the far range of the scene. In any case, the residual linear phase due to target's elevation can be also estimated and compensated by an interferometric processing, disposing of at least two ULAs displaced along $z$.\\

%We can gain insight on which of the two effects is more relevant by an example. Let us assume a constant nominal velocity $v_x$ along $x$ and a target placed in the direction of motion ($\phi = 0$), for which the error due to elevation mismatch is maximum. We have then:
%
%\begin{equation}
%    \Delta \psi^z(\mathbf{x}, \tau) = \frac{4\pi}{\lambda}\frac{v_x \tau }{r \tan\theta}\Delta z. 
%\end{equation}
%
%When the radar is mounted in the front of the car (frontal looking configuration), $\theta \approx 90$ deg for all the FoV. For a system working at $77$ GHz ($\lambda \approx 4$ mm), travelling at $v_x = 15$ m/s, a target at slant range $r = 20$ m is seen by an incidence angle $\theta \approx 87$ deg. An elevation error of $\Delta z = 1$ m maps into a residual Doppler frequency of $19$ Hz. The same residual Doppler frequency is induced by a velocity error along the direction of motion of $\Delta v_x = 3.9$ cm/s only. Therefore, even in the worst possible condition (the target placed along the direction of motion) the elevation mismatch $\Delta z$ will poorly affect the residual phase.

\subsubsection{Error in the angular localization of GCP}
\begin{figure}[!t]
    \centering
    \includegraphics[width=0.9\columnwidth]{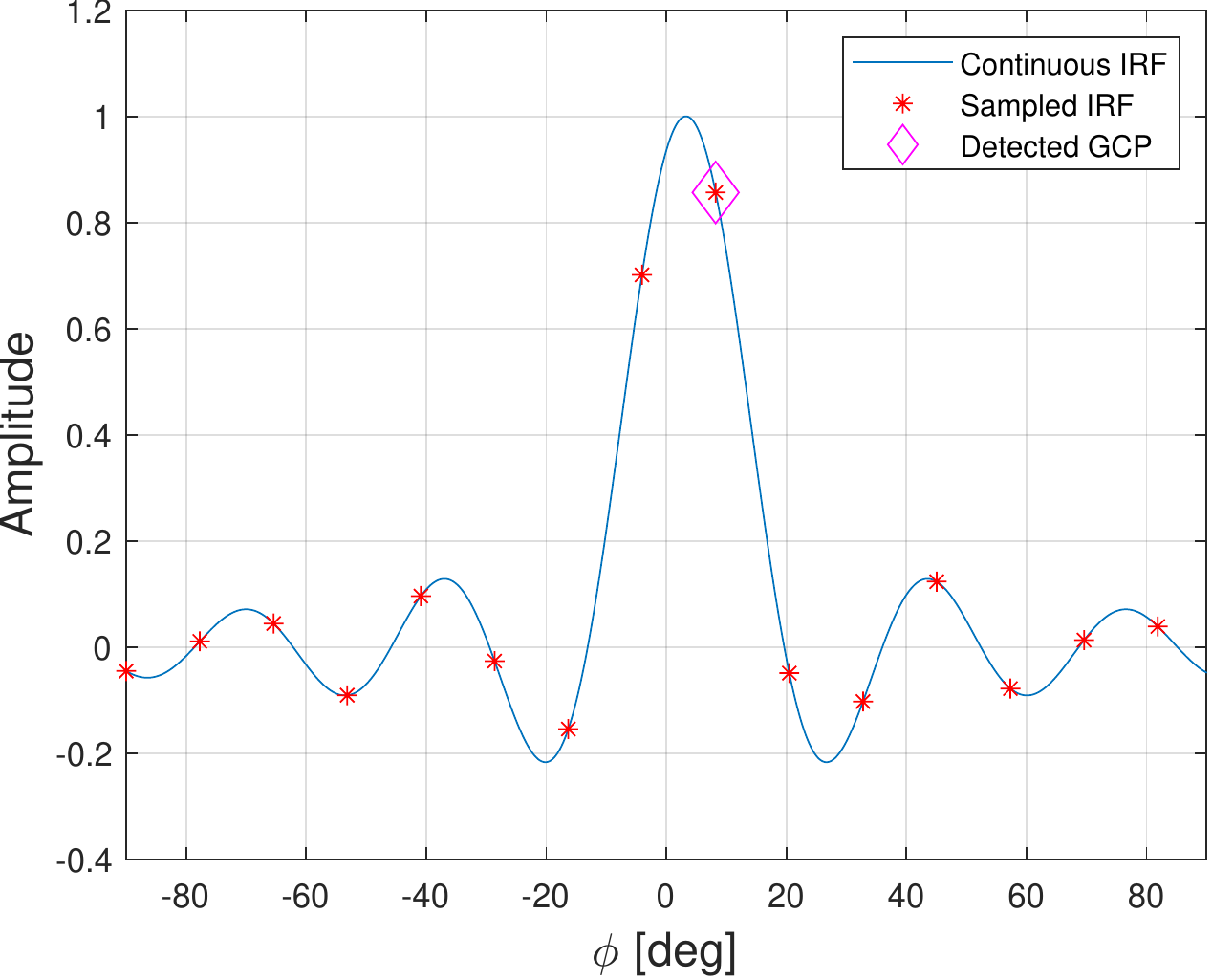}
    \caption{Example of angular detection error on a GCP. (blue) the continuous Impulse Response Function of the MIMO array. (red) Sampling position. (purple) the detected GCP is not exactly on the peak of the cardinal sine function.}
    \label{fig:angular_error}
\end{figure}

The same reasoning used for an error in the focusing height can be applied to an error in the angular localization of a GCP. Without velocity focusing height errors, the residual Doppler frequency is exactly zero over a given target. Due to the finite sampling of the MIMO images, however, it can happen that the peak of the Impulse Response Function (IRF) representing the target is not detected, as shown in Figure \ref{fig:angular_error}. The mis-detection could happen also due to the in-avoidable presence of noise. In blue, the continuous IRF with an angular resolution given by the physical length of the array, the sampling positions being depicted in red while the detected GCP in purple. If the detected GCP is not at the peak of the cardinal sine function, a residual Doppler frequency is present.

\begin{figure}[!t]
    \centering
    \includegraphics[width=0.9\columnwidth]{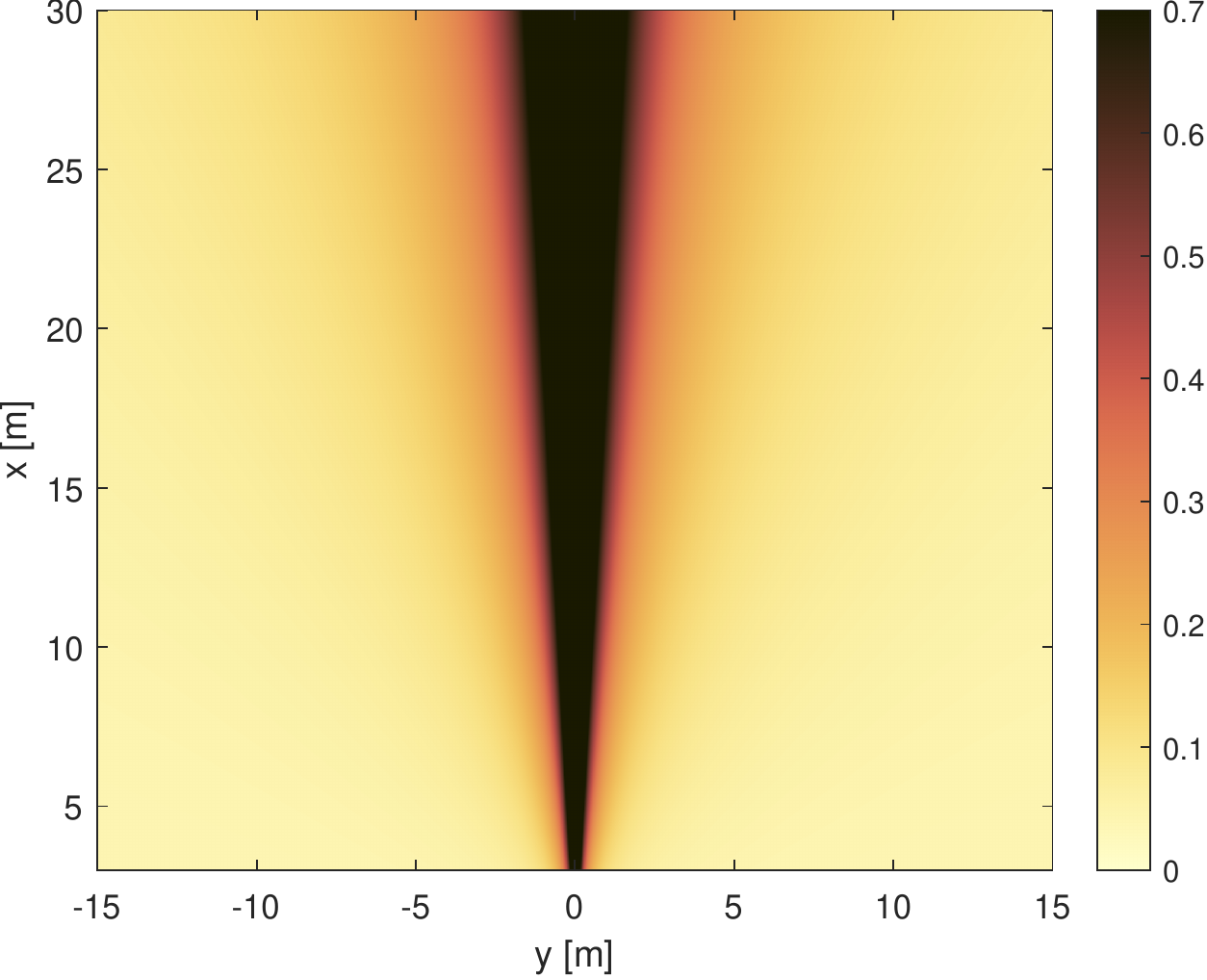}
    \caption{Maximum tolerable angular error in GCP detection. The nominal vehicle velocity  is  set  to  15  m/s  along $x$,  the  nominal  radar  height  over  ground  is 0.5 m, the maximum radar slant range is 30 m. The system is tolerable to angular errors in the direction of the motion.}
    \label{fig:maximum_angular_error}
\end{figure}

Let us call $\Delta \phi$ the error on the angular localization of a GCP. We have:
\begin{equation}
\label{eq:residual_phase_angular}
\begin{split}
        \Delta \psi^\phi(\mathbf{x}, \tau) &= \frac{\partial \psi(\mathbf{x}, \tau)}{\partial \phi}\Delta \phi = \left[\left(\mathbf{k}'_\phi(\mathbf{x})^\mathrm{T} \mathbf{v} \right) \Delta \phi\right]  \tau
\end{split}
\end{equation}
where $\mathbf{k}_\phi'(\mathbf{x})=(4\pi/\lambda)[-\sin\theta\sin\phi, \, \sin\theta\cos\phi,\, 0]^\mathrm{T}$ is the derivative of $\mathbf{k}(\mathbf{x})$ with respect to $\phi$. The residual angular Doppler frequency is then:
\begin{equation}
\label{eq:residual_doppler_angular}
\begin{split}
    \Delta \omega^\phi = \frac{\partial \psi(\mathbf{x}, \tau)}{\partial \phi}\Delta \phi &= \left(\mathbf{k}'_\phi(\mathbf{x})^\mathrm{T} \mathbf{v} \right) \Delta \phi = \frac{4\pi}{\lambda}v_\perp \Delta \phi.
\end{split}
\end{equation}
Again, the maximum tolerable angular error is the one that will generate a Doppler frequency corresponding to $\Delta v_r^\text{max}$.
In Figure \ref{fig:maximum_angular_error}, the maximum tolerable angular error $\Delta \phi$ is depicted for every pixel in the field of view. The car is supposed to travel along $x$ at $15$ m/s. In this case, $v_\perp$ is $\approx 0$ for the pixels exactly in front of the car, thus allows for a larger $\Delta \phi$. The worst case is for $\phi \approx 90$ deg, where the accuracy in GCP detection must be the maximum. GCP detection accuracy can be refined with a parabolic interpolation to reduce unwanted phase effects.\\
It is important to highlight that, with a sufficient number of GCPs, the angular error becomes irrelevant since the errors will be both positive and negative and the average will approach zero. The same is not true for elevation error where a significant bias towards positive or negative errors may be present.\\

Once the GCPs have been detected, a frequency estimation is performed through a Fast Fourier Transform (FFT) of the phase of each GCP, and the position of the peak in the frequency domain is extracted to form $\Delta \boldsymbol{\omega}$ in \eqref{eq:linear_system}. 
It is possible that a moving target (bike, another vehicle, pedestrian, etc.) is detected as a GCP, preventing a correct residual motion estimation. It is mandatory to discard outliers before the inversion of \eqref{eq:linear_system}, by imposing a threshold on the maximum frequency that is possible to find on a given GCP. As a rule of thumb, the value of the threshold can be derived from the accuracy of the navigation data: if the the nominal accuracy is, for instance, $20$ cm/s, it is unlikely to find a GCP with a residual Doppler frequency much higher than the one corresponding to $20$ cm/s. It is worth to stress that, the more accurate the navigation data, the more robust is the outlier rejection and, consequently, the better the performance of the whole procedure. 
The linear system \eqref{eq:linear_system} can therefore be solved using the Weighted Least Square (WLS) method:
\begin{equation}
\label{eq:residual_velocities}
    \widehat{\Delta\mathbf{v}} = (\mathbf{K}^\mathrm{T}\mathbf{W}\mathbf{K})^{-1}\mathbf{K}^\mathrm{T}\mathbf{W}\Delta\boldsymbol{\omega}
\end{equation}
where $\mathbf{W}$ is a proper weighting matrix. Each GCP can be weighted according to some specific figure of merit such as the amplitude of the GCP or the prominence of the peak in the frequency domain (i.e., how much that GCP shows a sinusoidal behavior in the frequency domain). It is worth remarking that the matrix inversion \eqref{eq:residual_velocities} might be unstable. In a typical automotive environment, the radar is mounted close to the road, thus $\theta \approx 90$ deg for every GCP. The direct consequence is that the residual radial velocity vector of any given GCP has components only in the $x,y$ plane and not in the $z$ direction, therefore a residual velocity along $z$ has no possibility to be detected with satisfactory accuracy. It is even more straightforward from \eqref{eq:linear_phase}: if $\Delta v_z \neq 0$ there is no consequence on the residual phase if $\theta = 90$ deg. The solution can be to avoid the estimation of $\Delta v_z$ by removing the last column of $\mathbf{K}$ and the last row of $\Delta\mathbf{v}$.
The quality of the velocity estimate $\widehat{\Delta}\mathbf{v}$ can be assessed by its covariance matrix, that is:
% % la covarianza è il CRB solo se il rumore è gaussiano bianco, qui potrebbero romperci le scatole perchè il rumore in realtà è il rumore sulla stima di una frequenza... che rumore è quello? bianco? rosa? gaussiano? DDP uniforme?... meglio mettere solo covariance function.
\begin{equation}
\label{eq:covariance_estimate}
    \mathbf{C}_{\widehat{\Delta\mathbf{v}}} = (\mathbf{K}^\mathrm{T}\mathbf{W}\mathbf{K})^{-1}\sigma_n^2.
\end{equation}
Notice that, in practice, the exact value of $\sigma_n^2$ is unknown, but with a sufficient number of GCP it can be roughly estimated from the residual of \eqref{eq:linear_system}, providing some realistic values for the accuracy of the velocities' estimates. Equation \eqref{eq:covariance_estimate} can provide also some insights on how to choose GCP in the scene. For the sake of simplicity, we assume equal unitary weights for each GCP ($\mathbf{W}=\mathbf{I}$), we normalize the noise power ($\sigma_n^2 = 1$) and we consider a 2D geometry ($\theta = 90$ deg). In this scenario we have:
\begin{equation}
    \mathbf{C}_{\widehat{\Delta\mathbf{v}}} = \frac{1}{\lvert\mathbf{K}^\mathrm{T}\mathbf{K}\rvert}
    \begin{bmatrix}
    \mathbf{k}_y^\mathrm{T}\mathbf{k}_y & -\mathbf{k}_x^\mathrm{T}\mathbf{k}_y \\
    -\mathbf{k}_y^\mathrm{T}\mathbf{k}_x & \mathbf{k}_x^\mathrm{T}\mathbf{k}_x
    \end{bmatrix}
\end{equation}
where $\mathbf{k}_x = [k_x^0 , k_x^1 , \dots, k_x^P]^\mathrm{T}$,  $\mathbf{k}_y = [k_y^0 , k_y^1 , \dots, k_y^P]^\mathrm{T}$, and $\lvert\cdot\rvert$ denotes the determinant of a matrix. 
If we choose all the GCPs closely spaced in front of the car (i.e., $\phi\rightarrow 0$ deg), the estimation of $\Delta v_y$ will be much more unreliable than the estimate of $\Delta v_x$ since $\mathbf{k}_x^\mathrm{T}\mathbf{k}_x \gg \mathbf{k}_y^\mathrm{T}\mathbf{k}_y$. The same reasoning is valid for GCPs closely spaced at the side of the car (i.e., $\phi\rightarrow 90$ deg): the only reliable estimate is on $\Delta v_y$. The extreme case is when all the GCP are closely grouped together: in this case the matrix $\mathbf{K}$ is close to be singular and the estimates of the residual velocities are useless.
Notice that the residual velocities estimated in \eqref{eq:residual_velocities} can be also used to improve the ego-motion estimation of the vehicle. In this case it is sufficient to integrate the residual velocities to obtain the residual trajectory and then compensate the error in the original trajectory provided by the navigation unit.

\subsection{SAR image formation}
Once the residual velocities have been found, it is possible to compute the forward problem for each pixel $\mathbf{x}$ in the scene and for each $\tau$ (i.e., each one of the $M$ low resolution MIMO images). This leads to a set of estimated TPS:
\begin{equation}
    \widehat{\Delta\psi}(\mathbf{x},\tau) = \left(\mathbf{k}(\mathbf{x})^\mathrm{T} \widehat{\Delta\mathbf{v}}\right)\tau
\end{equation}
Each low resolution image is phase-compensated using the estimated TPS:
\begin{equation}
    \widehat{I}_m(\mathbf{x},\tau) = I_m(\mathbf{x},\tau)\exp\left\{-j\widehat{\Delta\psi}(\mathbf{x},\tau)\right\}.
\end{equation}
and then coherently summed to obtain the final high-resolution SAR image
\begin{equation}
\label{eq:final_MIMO_sum}
    I(\mathbf{x}) = \sum_{\tau \in T}\widehat{I}_m(\mathbf{x},\tau).
\end{equation}
The final SAR image is now properly focused, localized and ready to be used in safety critical systems such as advanced autonomous driving systems.

\begin{figure*}[!t]
    \centering
    \subfloat[Equipment]{\includegraphics[width=\columnwidth]{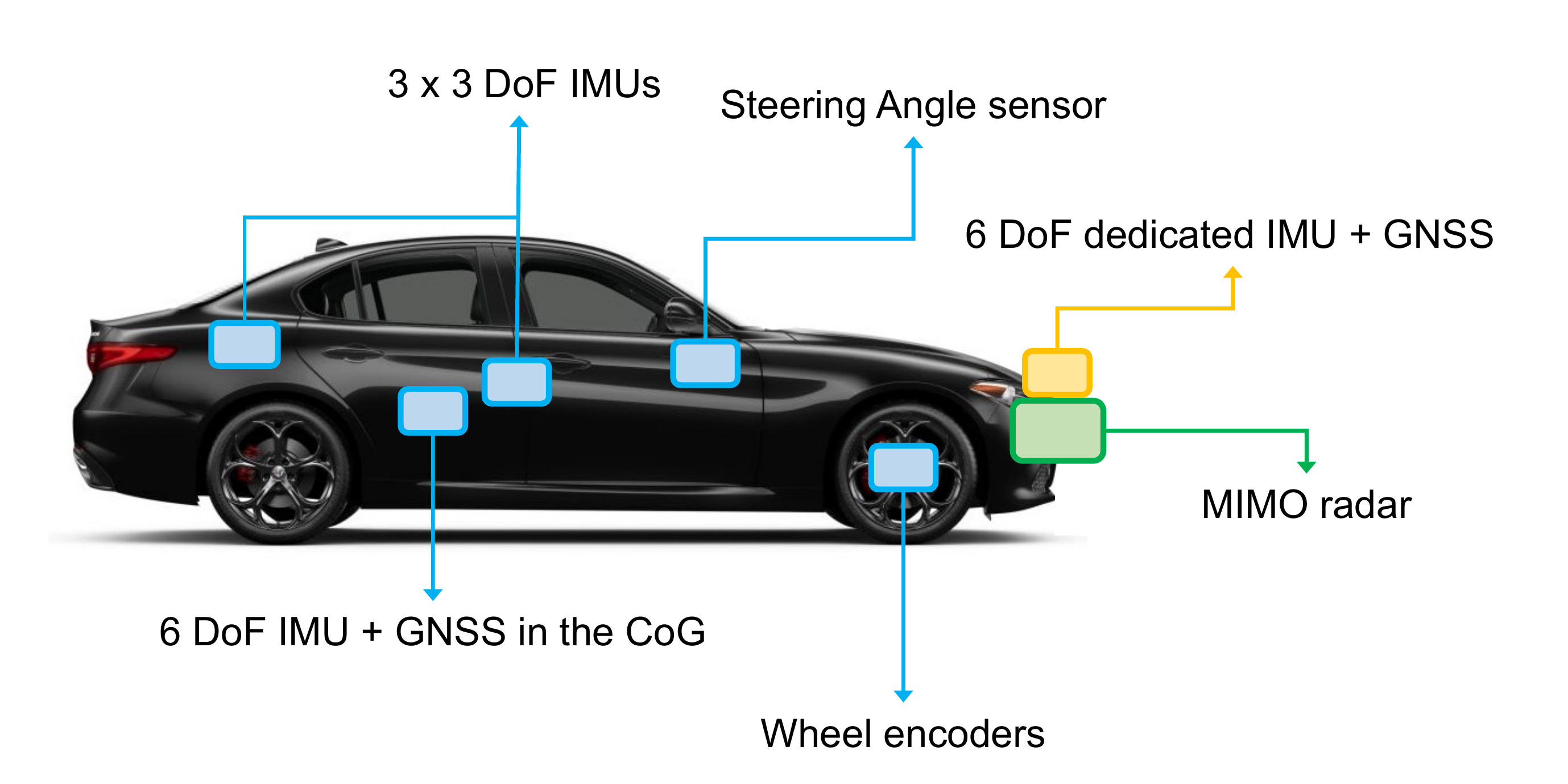}\label{subfig:equipment}}
    \subfloat[Camera]{\includegraphics[width=\columnwidth]{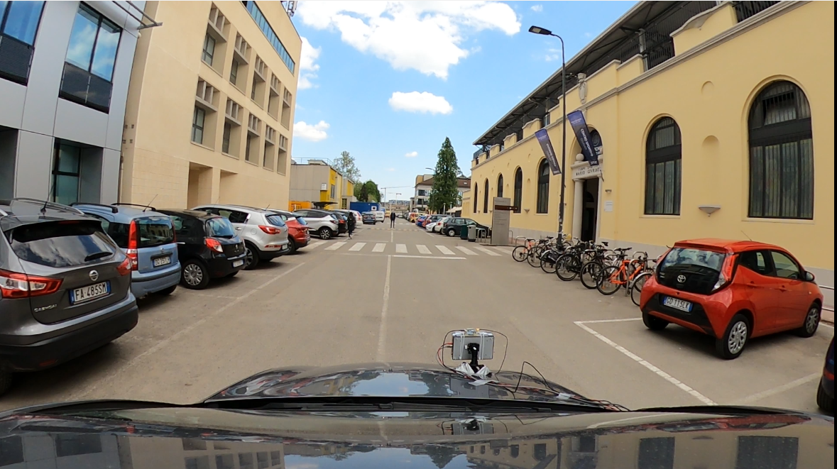}\label{subfig:gopro_road}}
    \caption{(\ref{subfig:equipment}) Vehicle navigation equipment; (\ref{subfig:gopro_road}) optical image of the field of view gathered by a camera mounted on to the vehicle.}
    \label{fig:equipment+gopro}
\end{figure*}

%%%%%%%%%%%%% RESULTS %%%%%%%%%%%%%%%%%%%%
\section{Results with real data}\label{sec:results}

\begin{table}[!t]
% increase table row spacing, adjust to taste
\renewcommand{\arraystretch}{1.3}
%if using array.sty, it might be a good idea to tweak the value of
%\extrarowheight as needed to properly center the text within the cells
\caption{System parameters used in the campaign}
\label{table:system_parameters}
\centering
% Some packages, such as MDW tools, offer better commands for making tables
% than the plain LaTeX2e tabular which is used here.
\begin{tabular}{|c||c|}
\hline
\textbf{Parameter} & \textbf{Value}\\
\hline
Carrier frequency ($f_c$) & 77 GHz \\
Bandwidth ($B$) & 3 GHz \\
Pulse length ($T_p$) & 55 $\mu$s \\
PRI & 1 ms\\
Active TX channels & 2 \\
Active RX channels & 4 \\
Maximum range & 27 m \\
Geometry (Mode) & forward looking \\
\hline
\end{tabular}
\end{table}

To validate the proposed technique, we carried out an acquisition campaign using a fully equipped vehicle. The radar system is a proprietary ScanBrick\textsuperscript{\textregistered} by Aresys\textsuperscript{\textregistered} and it is based on the Texas Instruments AWR1243 Single-Chip 77- and 79-GHz FMCW transceiver \cite{noauthor_awr1243_nodate}. The maximum available bandwidth is $4$ GHz for a range resolution up to $3.75$ cm. The mounting position of the radar on the vehicle is precisely known. The entire radar equipment is based on standard automotive hardware suitable for future mass-market production.\\
The radar is mounted in a forward looking geometry, thus the boresight of the MIMO array is pointing in the direction of motion, as it can be seen from Figure \ref{subfig:gopro_road}. We employed just 2 out of 3 Tx antennas and all the 4 Rx ones, leading to a virtual array of $N=8$ elements, spaced by $\lambda/4$. The angular resolution of the low-resolution images $I_m(\tau;\mathbf{x})$ is approximately $ 16$ deg. The transmitted signal has $3$ GHz of bandwidth leading to a range resolution of $5$ cm. All the system's parameters are summarized in Table \ref{table:system_parameters}.
Notice that in all the previous works in literature, the radar was mounted side looking. In fact, a SAR in forward looking geometry, without an array displaced along the direction orthogonal to the motion ($y$) would lead to a totally left/right ambiguous SAR image. The presence of a ULA with the elements displaced along $y$ helps to unambiguously reconstruct the image (as in our case study).

The car equipment (Figure \ref{subfig:equipment}) is complemented by navigation sensors to provide the estimated trajectory as input to the procedure. The on-board navigation equipment comprises: \textit{(i)} two internal 3 Degrees of Freedom (DoF) IMUs, measuring lateral and longitudinal acceleration, along with heading rate; \textit{(ii)} an occupant restraint controller, consisting in a 3 DoF IMU placed in the rear part of the car, measuring longitudinal and lateral acceleration as well as heading rate, the purpose is the airbag activation during a crash; \textit{(iii)} four wheel encoders, measuring the odometric velocity of each wheel; \textit{(iv)} a steering angle sensor at the frontal wheels; \textit{(v)} an on-radar IMU+GNSS sensor~\cite{InertialSense2020}. The sensors' data are fused with an Unscented Kalman Filter (UKF) approach described in our previous work~\cite{tagliaferri_navigation-aided_2021}.

The acquisition campaign has been carried out in a straight road with several targets in the FOV of the radar, as depicted in Figure \ref{subfig:gopro_road}, showing the environment image acquired by a camera mounted on top the vehicle. We selected a portion of the trajectory made by $M=200$ slow time samples and we processed the dataset with and without running the autofocus workflow. The nominal speed of the vehicle in the selected synthetic aperture was $25$ km/h, thus, for a PRI of $1$ ms, leads to an average synthetic aperture length $A_s\approx1.4$ m.

\begin{figure}[!t]
    \centering
    \includegraphics[width=0.9\columnwidth]{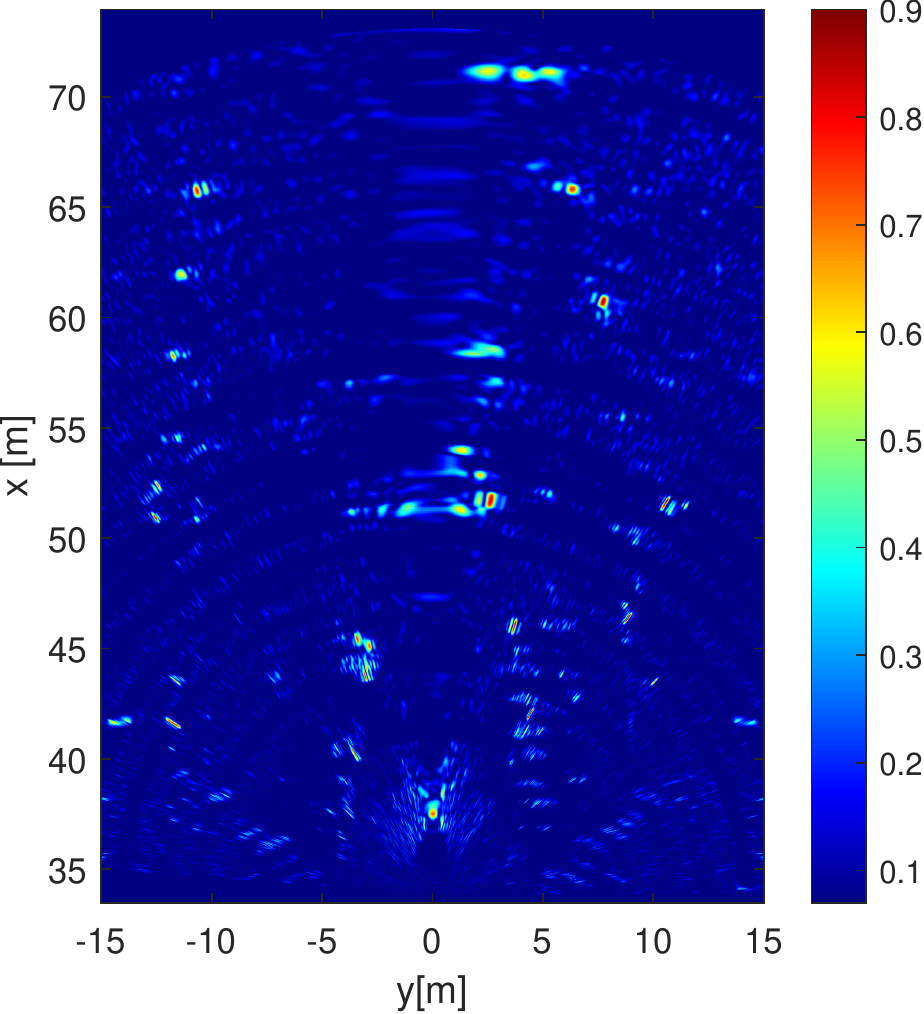}
    \caption{SAR intensity image without employing any autofocus procedure. The amplitudes are normalized and in linear scale.}
    \label{fig:no_autofocus}
\end{figure}

The result of the SAR processing without autofocus (only navigation-based MoCo) is depicted in Figure \ref{fig:no_autofocus}. It is interesting to notice how, in far range, the image is completely corrupted by an error on the estimated trajectory. The image seems to collapse inward (i.e., towards the line at $x=0$). Some details are still preserved and not totally defocused in near range, such as the bikes and the cars parked at $y=\pm 5$ m. Nevertheless, the localization accuracy of the targets might not sufficient for safety-critical autonomous driving systems.

\begin{figure}[!t]
    \centering
    \includegraphics[width=0.9\columnwidth]{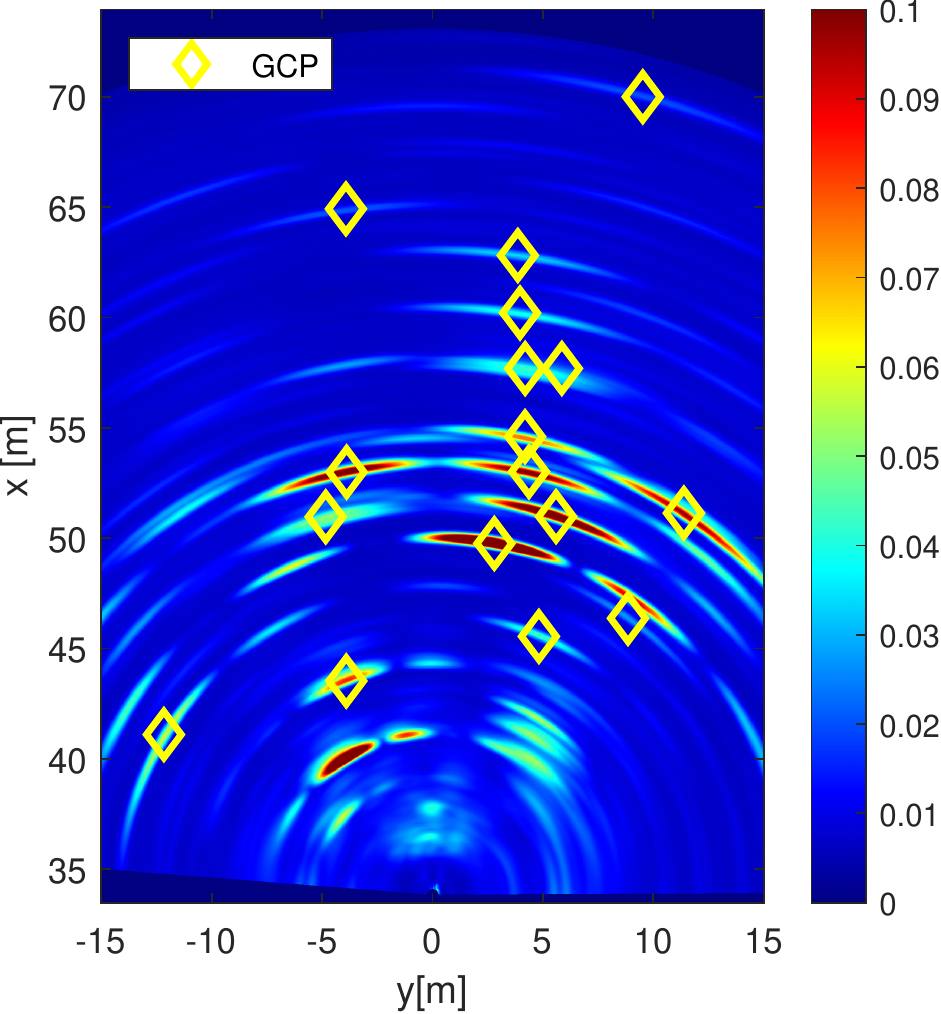}
    \caption{Incoherent mean of all the low resolution images. The selected anchors are depicted with yellow diamonds.}
    \label{fig:mimo_anchors}
\end{figure}

\begin{figure*}[!t]
    \centering
    \subfloat[]{\includegraphics[width=\columnwidth]{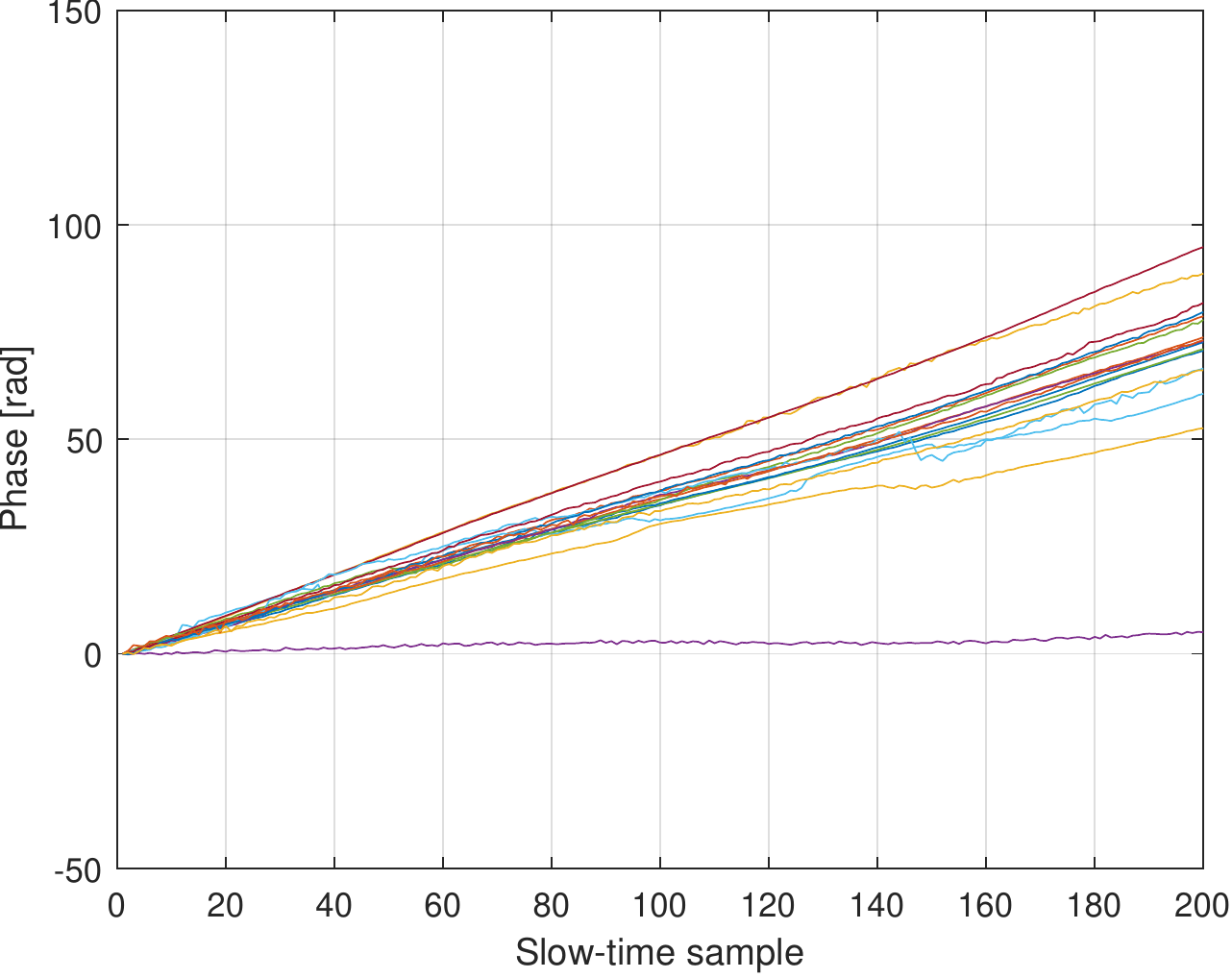}\label{subfig:unwrapped}}
    \subfloat[]{\includegraphics[width=0.98\columnwidth]{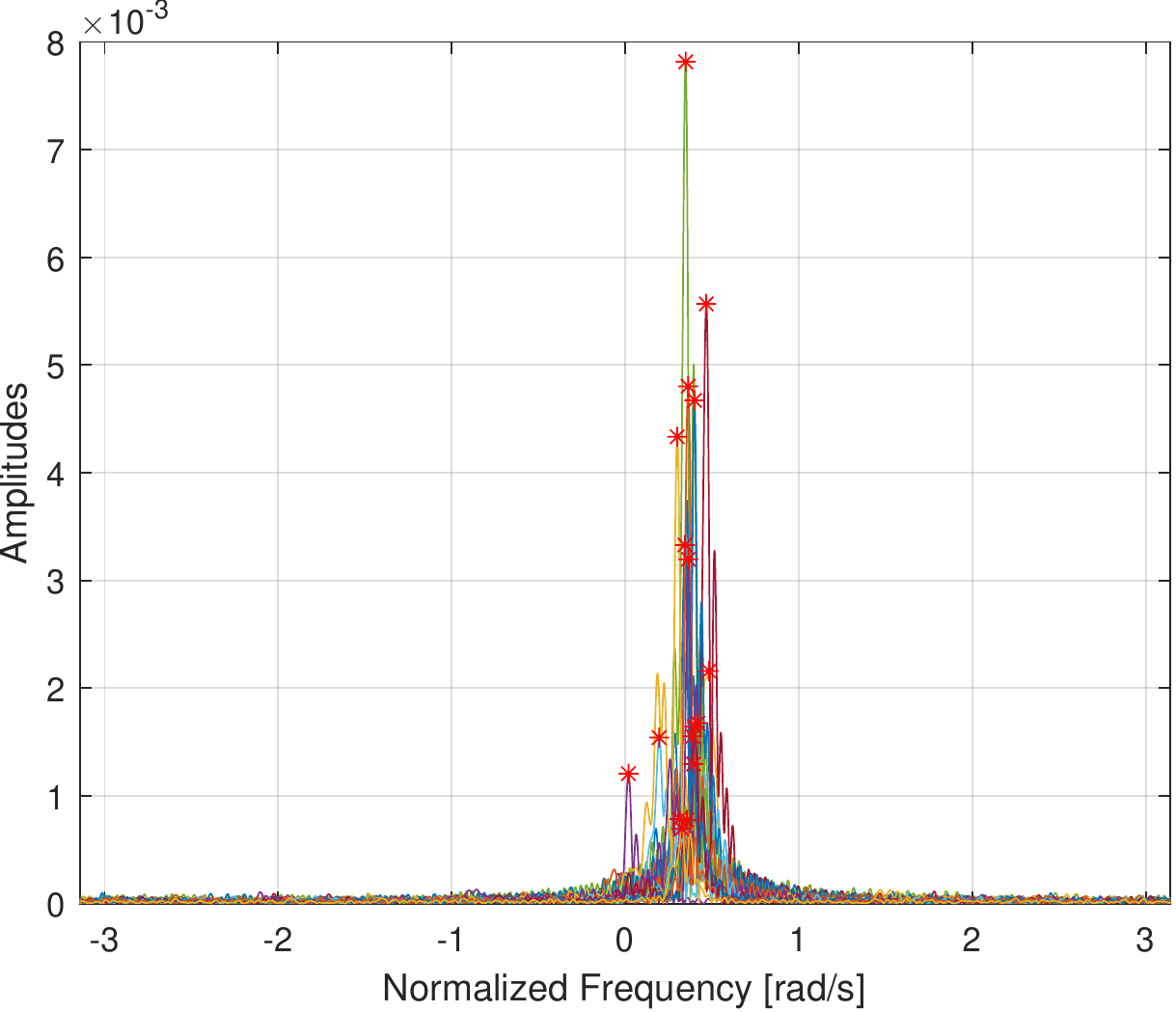}\label{subfig:frequency_analysis}}
    \caption{(\ref{subfig:unwrapped}) Unwrapped phase over the selected GCP: a residual phase is linear in slow time indicating the presence of a residual motion error; (\ref{subfig:frequency_analysis}) Result of the frequency estimation over the selected GCP: a linear phase represents a sinusoid. Notice how the cardinal sines functions are not centered in zero.}
    \label{fig:unwrapped+frequency}
\end{figure*}

The autofocus workflow starts from the detection of a set of GCPs from the incoherent average of all the low resolution images. In Figure \ref{fig:mimo_anchors}, the incoherent average is represented along with the detected GCPs (highlighted in red). The first observation is that, as expected, the spatial resolution of the MIMO image formed with the ULA is greatly lower than Figure \ref{fig:no_autofocus}). The second and most important observation is that the low-resolution images are not severely corrupted by trajectory error as the SAR one. While the cars parked in the scene at $y = -5$ m and $x \in [35,45]$ m are tilted inwards in Figure \ref{fig:no_autofocus}, they are correctly straight (but with lower resolution) in Figure \ref{fig:mimo_anchors}. 
%This is an expected behavior: the cars are parked straight (See Figure \ref{subfig:gopro_road}) but the coherent processing with a wrong trajectory generates a significant displacement in the reconstructed scene.
Over the detected GCPs, the residual Doppler frequency is then estimated through FFT. In Figure \ref{subfig:unwrapped}, the unwrapped residual phase over all the GCPs is depicted. The phase is linear for every GCP and the slope is proportional to the residual radial velocity of the car as seen by the position of the GCP. The result of the FFT is depicted in Figure \ref{subfig:frequency_analysis} and the red dots are the detected frequency peaks. The position of such peaks will form the observation vector $\Delta \boldsymbol{\omega}$ in \eqref{eq:linear_system}. The frequency resolution depends on the observation time, in our case the length of the synthetic aperture $T$. A trade off is now evident: longer synthetic apertures allows for higher spatial resolution and, from the autofocus perspective, a higher residual Doppler resolution. The price to be paid is the possibility of non-constant velocity errors in longer apertures (acceleration errors) and increased computational burden. It is also important to notice that Figure \ref{subfig:frequency_analysis} shows the \textit{sampled} version of the Doppler spectrum of a GCP. The sampling in the frequency domain can be made finer by zero-padding the time domain signal before the FFT. This guarantees that the position of the peak of the cardinal sine function is precisely detected.

The inverse problem \eqref{eq:residual_velocities} is solved for the detected GCPs leading to the residual velocities in Table \ref{table:residual_velocities}, reported with the theoretical accuracy. First of all, the estimated residual velocities are within the confidence bound of the employed navigation sensors~\cite{tagliaferri_navigation-aided_2021}. As expected, the error is higher in the direction of motion: in a forward looking geometry, all the GCPs are distributed in front of the car, thus higher accuracy is expected in this direction. On the other hand, the error in the direction orthogonal to the motion is much lower and estimated with more uncertainty. 

\begin{figure}[!t]
    \centering
    \includegraphics[width=0.9\columnwidth]{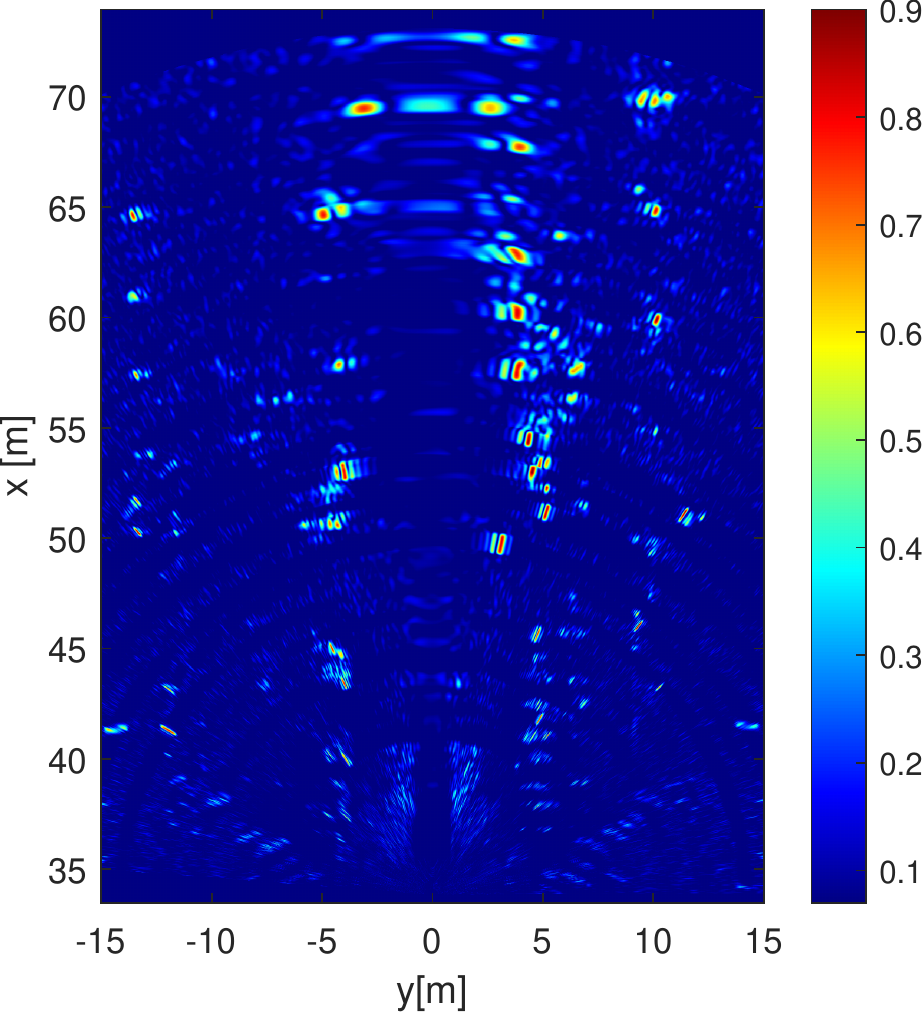}
    \caption{SAR intensity image after employing the proposed autofocus procedure. The amplitudes are normalized and in linear scale.}
    \label{fig:autofocus}
\end{figure}

\begin{figure*}[!t]
    \centering
    \includegraphics[width=0.9\textwidth]{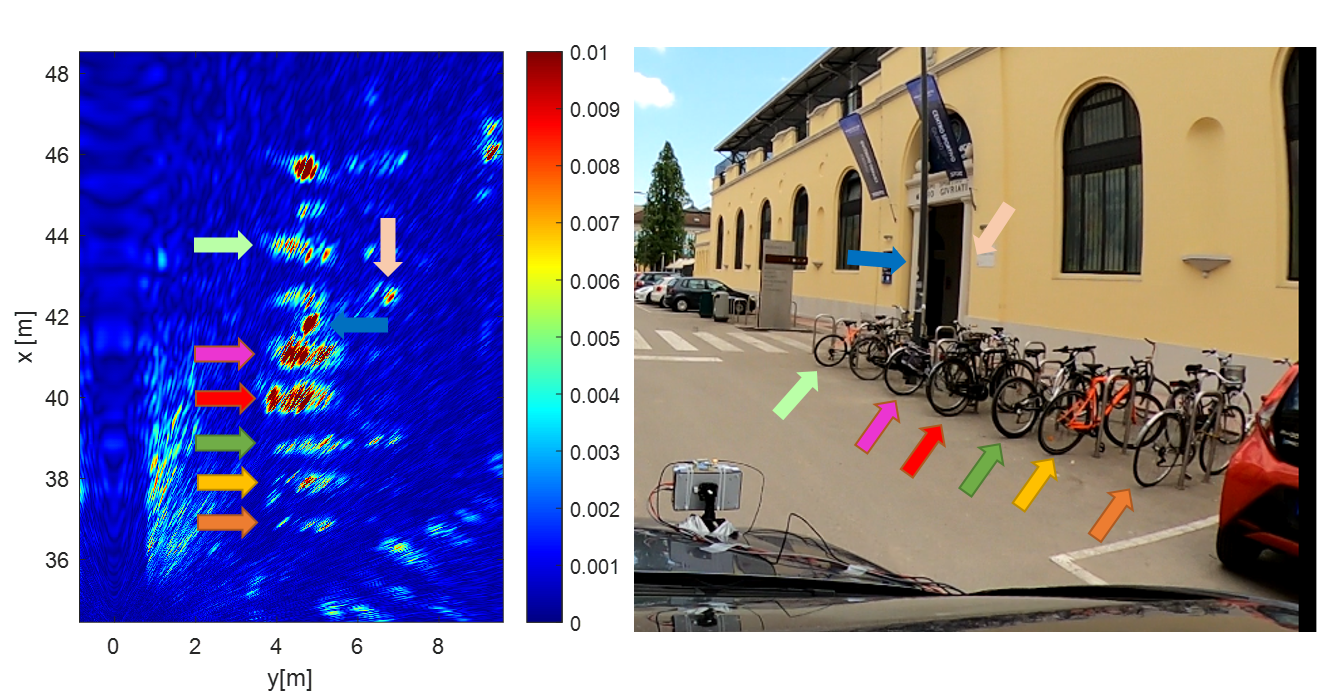}
    \caption{(left) SAR intensity image after autofocusing. A few details are recognizable from the figure on the right (optical image). Five bicycles (orange, yellow, green, red and purple arrows), a lighting pole (blue arrow), a marble column (pink arrow) and another bicycle (light green arrow). The amplitude is not normalized.}
    \label{fig:details_1}
\end{figure*}

\begin{figure*}[!t]
    \centering
    \includegraphics[width=0.9\textwidth]{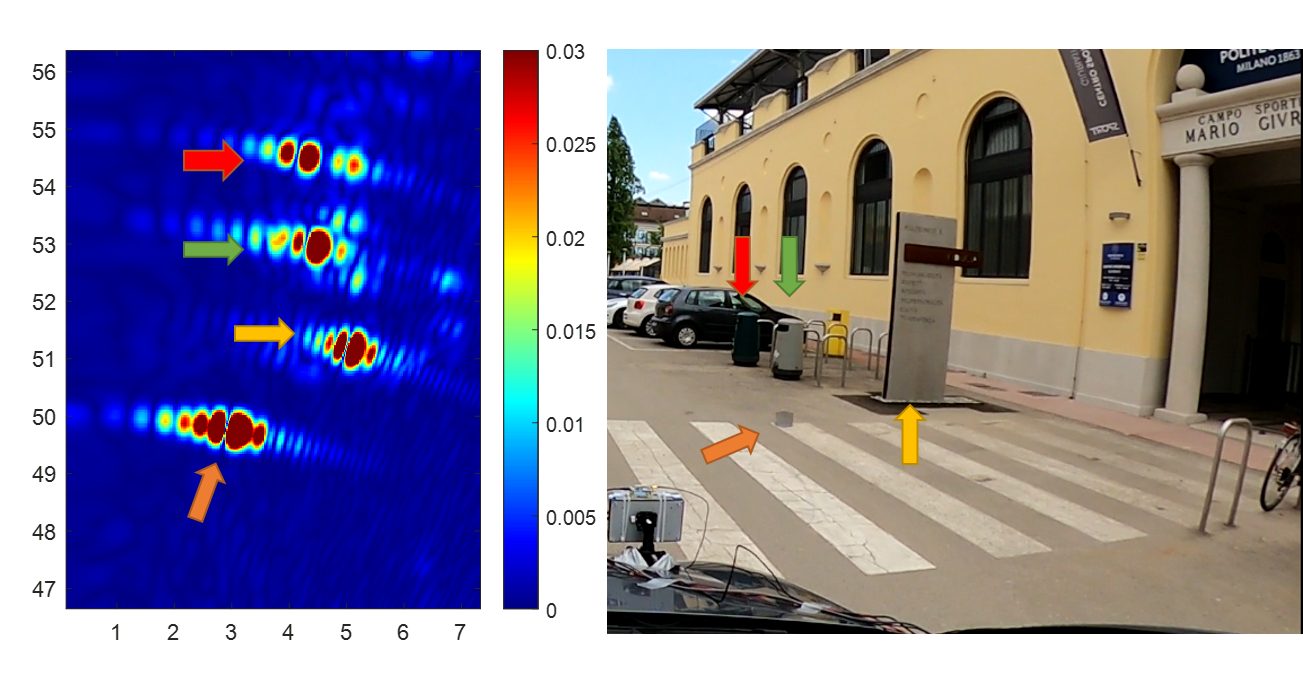}
    \caption{(left) SAR intensity image after autofocusing. Some other details are recognizable from the figure on the right (optical image). One corner reflector (orange arrow), a marble stele (yellow arrow) and two garbage cans (red and green arrows). The amplitude is not normalized.}
    \label{fig:details_2}
\end{figure*}

Once the residual velocities are estimated, each low resolution image is TPS-compensated and the coherent average forms the final SAR image as from \eqref{eq:final_MIMO_sum}. The image is represented in Figure \ref{fig:autofocus}. While Figure \ref{fig:no_autofocus} reports a collapsed scene towards the center of the image, now the profile given by the parked cars is correctly straight also in far range. A few details are depicted in Figure \ref{fig:details_1}. On the left of the figure, a zoomed version is presented. In this portion of the image it is possible to distinguish the five bicycles parked at the right of the road (orange, yellow, green, red and purple arrows), the lighting pole (blue arrow), the marble column (pink arrow) and the next bicycle after the lighting pole (light green arrow). On the right, the optical image is depicted for comparison. Moving forward in the trajectory (Figure \ref{fig:details_2}) other details appear, such as the corner reflector placed on the ground (orange arrow), the marble stele (yellow arrow) and the two garbage cans (red and green arrows).
The SAR image can now be used as an input product along with LiDAR or cameras in advanced autonomous driving systems. It is interesting to notice that the proposed algorithm is fast and parallelizable, since it requires just the computation of a set of Fourier transforms. This characteristic makes it suitable for a real time implementation in automotive scenarios.

\begin{table}[!t]
% increase table row spacing, adjust to taste
\renewcommand{\arraystretch}{1.3}
%if using array.sty, it might be a good idea to tweak the value of
%\extrarowheight as needed to properly center the text within the cells
\caption{Residual velocity estimated by autofocus and related accuracy}
\label{table:residual_velocities}
\centering
% Some packages, such as MDW tools, offer better commands for making tables
% than the plain LaTeX2e tabular which is used here.
\begin{tabular}{|c||c||c|}
\hline
\textbf{Parameter} & \textbf{Estimate (cm/s)} & \textbf{Accuracy (cm/s)}\\
\hline
$\Delta v_x$ & 22.78 & 1.27\\
$\Delta v_y$ & 1.07 & 2.24\\
\hline
\end{tabular}
\end{table}

\section{Conclusion}
\label{sec:conclusions}
The generation of SAR images requires a precise knowledge of the trajectory of the moving platform. For automotive applications and synthetic apertures of tens of centimeters to few meters length, velocity estimation errors from inaccurate navigation data are the major source of SAR image quality degradation, causing defocusing and targets' mis-localization. The higher the carrier frequency and the synthetic aperture length, the higher the accuracy needed on velocity estimation, with maximum tolerable errors as low as $1$ cm/s. In these cases, inexpensive automotive-legacy navigation systems based on in-car sensors are not accurate enough, as the accuracy is typically not lower than $10$ cm/s.

This paper analytically derives the effect of typical residual motion estimation errors on automotive SAR focusing, setting the theoretical required accuracy on velocity estimation to avoid image degradation. In addition, we propose a complete residual motion estimation and compensation workflow based on both navigation data \textit{and} a set of low-resolution images generated by a physical or virtual radar array mounted on the car. First, a frequency analysis is carried out over a set of static GCPs detected in the low-resolution images, retrieving the residual vehicle's velocity. The latter is then used to a phase-compensate the low-resolution images, obtaining well-focused SAR images. We evaluate the impact of errors in both the estimated target's height and angular sampling of low-resolution radar images, providing guidelines on how to properly choose GCPs. As a rule of thumb, it is recommended to choose a sparse set of GCP in far range, allowing for a better estimate of both longitudinal and transversal components of the residual velocity and minimizing the effects of a wrong focusing height of the GCPs. Moreover, the more accurate are navigation data, the more robust is the selection of GCPs against outliers, justifying the joint usage of navigation and radar data.

The entire workflow is validated using a real dataset acquired using a forward-looking MIMO radar working in W-band mounted on a vehicle moving in an open road. The proposed workflow has proven to be able to estimate centimetric velocity errors and then to correctly recover the SAR image. 

% if have a single appendix:
%\appendix[Proof of the Zonklar Equations]
% or
%\appendix  % for no appendix heading
% do not use \section anymore after \appendix, only \section*
% is possibly needed

% use appendices with more than one appendix
% then use \section to start each appendix
% you must declare a \section before using any
% \subsection or using \label (\appendices by itself
% starts a section numbered zero.)
%

%\appendices
%\section{Proof of the First Zonklar Equation}
%Appendix one text goes here.

% you can choose not to have a title for an appendix
% if you want by leaving the argument blank
%\section{}
%Appendix two text goes here.

% use section* for acknowledgment
\section*{Acknowledgment}

The research has been carried out in the framework of the Huawei-Politecnico di Milano Joint Research Lab on automotive SAR. The Authors want to acknowledge Dr. Paolo Falcone from Aresys for the cooperation and support in the data acquisition campaign.

% Can use something like this to put references on a page
% by themselves when using endfloat and the captionsoff option.
\ifCLASSOPTIONcaptionsoff
  \newpage
\fi

% trigger a \newpage just before the given reference
% number - used to balance the columns on the last page
% adjust value as needed - may need to be readjusted if
% the document is modified later
%\IEEEtriggeratref{8}
% The "triggered" command can be changed if desired:
%\IEEEtriggercmd{\enlargethispage{-5in}}

% references section

% can use a bibliography generated by BibTeX as a .bbl file
% BibTeX documentation can be easily obtained at:
% http://mirror.ctan.org/biblio/bibtex/contrib/doc/
% The IEEEtran BibTeX style support page is at:
% http://www.michaelshell.org/tex/ieeetran/bibtex/
%\bibliographystyle{IEEEtran}
% argument is your BibTeX string definitions and bibliography database(s)
%\bibliography{IEEEabrv,../bib/references.bib}
%
% <OR> manually copy in the resultant .bbl file
% set second argument of \begin to the number of references
% (used to reserve space for the reference number labels box)
%\begin{thebibliography}{1}

%\bibitem{IEEEhowto:kopka}
%H.~Kopka and P.~W. Daly, \emph{A Guide to \LaTeX}, 3rd~ed.\hskip 1em plus
%  0.5em minus 0.4em\relax Harlow, England: Addison-Wesley, 1999.

%\end{thebibliography}

\bibliographystyle{IEEEtran}
%\bibliography{Bibliography.bib,references.bib}
\bibliography{references.bib}

\end{document}